\documentclass[
aps,prd,
nofootinbib,
superscriptaddress,
showpacs,
tightenlines,
]{revtex4}
\usepackage{amsmath}
\usepackage{amssymb}
\usepackage{bm}
\usepackage{color,graphicx}

\begin{document}

\title{Flavor dependence of chiral odd generalized parton distributions and the tensor charge from the analysis of combined $\pi^0$ and $\eta$ exclusive electroproduction data}

\author{Gary R.~Goldstein} 
\email{gary.goldstein@tufts.edu}
\affiliation{Department of Physics and Astronomy, Tufts University, Medford, MA 02155 USA.}

\author{J. Osvaldo Gonzalez Hernandez} 
\email{jog4m@virginia.edu}
\affiliation{Department of Physics, University of Virginia, Charlottesville, VA 22904, USA.}

\author{Simonetta Liuti} 
\email{sl4y@virginia.edu}
\affiliation{Department of Physics, University of Virginia, Charlottesville, VA 22904, USA.}
\affiliation{Laboratori Nazionali di Frascati, INFN, Frascati, Italy}


\begin{abstract}
Using a  physically motivated parameterization based on the reggeized diquark model we perform a flavor separation of the chiral-odd
generalized parton distributions obtained from both $\pi^0$ and $\eta$ exclusive electroproduction. In our approach we exploit a connection between the chiral-even  and chiral-odd reduced helicity amplitudes using Parity transformations 
that are relevant for a class of models that
includes two-component models. 
We compare our results for $\eta$ production to the previously obtained $\pi^0$ results, and we make predictions for the transverse single spin asymmetry components which will be measured within the Jefferson Lab 12 GeV program.
\end{abstract}

\maketitle

\baselineskip 3.0ex
\section{Introduction}
\label{sec:1}
\begin{figure}
\includegraphics[width=9.cm]{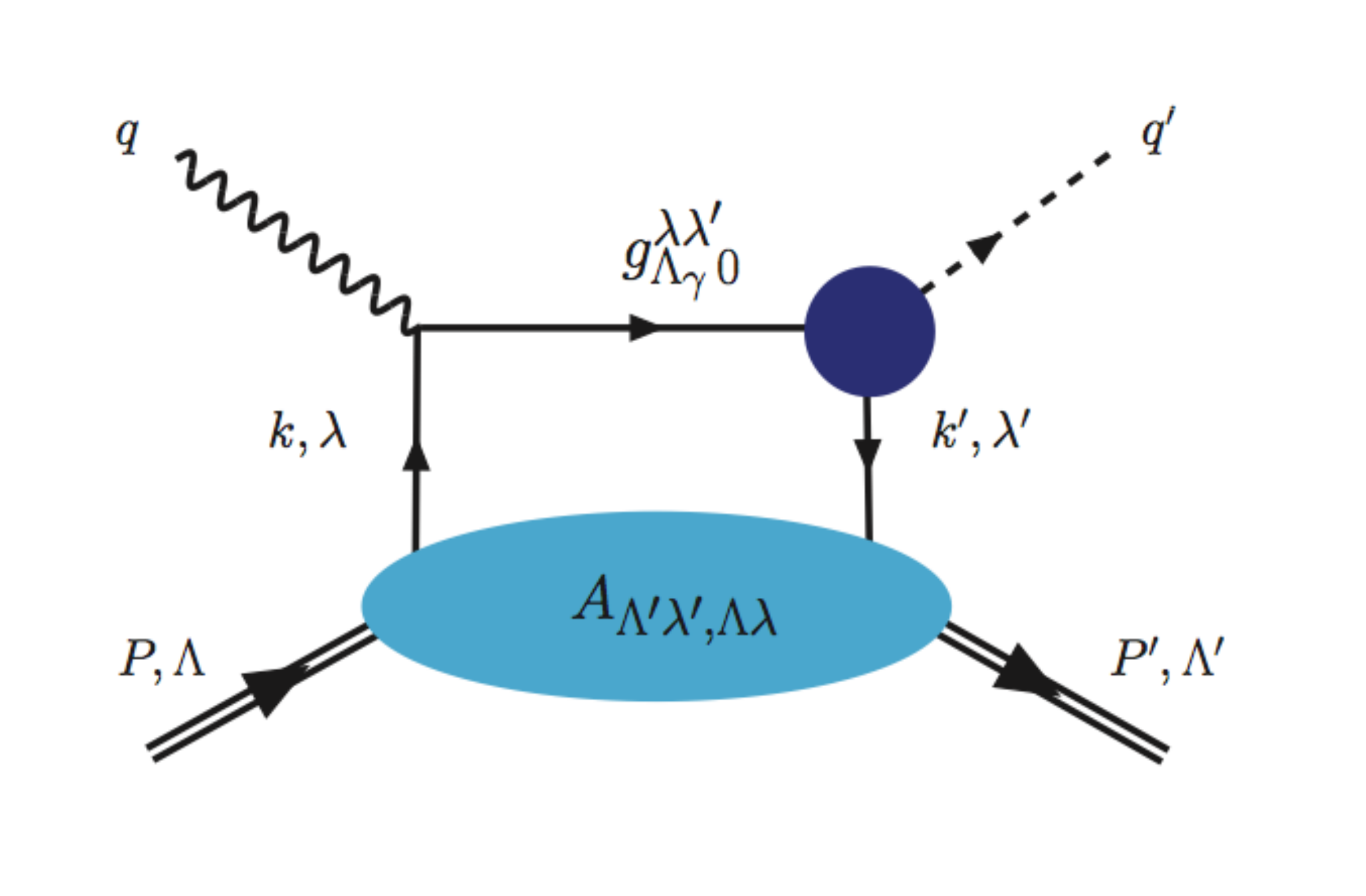}
\caption{Leading order amplitude for DV$\pi^0(\eta)$P, $\gamma^* + P \rightarrow M +P^\prime$. Crossed diagrams are not shown in the figure.}
\label{fig1}
\end{figure}

Deeply virtual exclusive pseudoscalar meson production  can be described within QCD factorization,
through the convolution of Generalized Parton Distributions (GPDs) and hard scattering amplitudes (Fig.\ref{fig1}).  
Although a full proof of factorization theorems was given only  for longitudinal photon polarization \cite{ColFraStr},  
in a series of papers \cite{AGL,GGL_pi0short,GGL_pi0} we showed that transverse photon polarization amplitudes can give substantial contributions  even if they appear at the next to leading twist
through the  $\pi^0$ coupling $\propto \gamma_5$, in the $\gamma^* p \rightarrow \pi^0 p'$ reaction. 
Because of the $\gamma_5$ coupling, the transverse polarization amplitudes contain convolutions of the four chiral-odd GPDs namely,
$H_T, E_T, \widetilde{H}_T, \widetilde{E}_T$ \cite{Ji_odd,Diehl_odd}.  

The chiral odd GPDs acquire a specific physical meaning, allowing us to explore different transverse spin configurations in the proton,  when written in terms of the quark-proton transversity amplitudes, $A^{T_{Y(X)}}_{\Lambda' \lambda', \Lambda \lambda}$, where $T_{Y(X)}$ represents the spin component along the $y(x)$-axis. Sensible information is obtained when the GPDs are proportional to linear combinations of amplitudes that are diagonal in a given basis, meaning that the spin projections are the same on the LHS and RHS of Fig.\ref{fig1}. This allows us to associate each  GPD  with parton distribution functions carrying the same spin information, bearing in mind that even if a connection between spin configurations is established, GPDs are related to amplitudes {\it i.e}. they are not probabilities in  momentum space (the quarks carry different momenta on the LHS and on the RHS of Fig.\ref{fig1}).
The connection between GPDs, Transverse Momentum Distributions (TMDs) both through their common ``parent" distributions, the Generalized TMDs (GTMDs) has been explored in \cite{Metz1,Metz2}, and in transverse coordinate space by Burkardt \cite{Bur1,Bur2}.  

In particular, one can see that  $H_T$ is the off forward generalization of the proton's transversity structure functions, $h_1$, or the probability of finding a  transversely polarized quark inside a transversely polarized proton.
$h_1$  and its integral over $x_{Bj}$, the tensor charge, $\delta$, have notoriously been elusive quantities to extract from experiment. Being chirally odd, $h_1$ can be measured in either Semi Inclusive Deep Inelastic Scattering (SIDIS) or in the Drell Yan process in conjunction with another chiral odd partner. The tensor charge's flavor dependence was obtained only relatively recently from model dependent analyses of SIDIS single hadron and dihadron production processes in the few GeV region, and for $x_{Bj} \gtrsim 0.06$ \cite{Anselmino,Courtoy}. 

The combination $2 \widetilde{H}_T + E_T$ has a helicity structure similar to one for the Boer Mulders function $h_{1}^\perp$ \cite{BoerMul}, the transversely polarized quark distribution in an unpolarized proton. Although one can show that the two distributions are generated from the same GTMD \cite{Metz1}, they have an important phase difference. As a result, $2 \widetilde{H}_T + E_T$, which involves a single spin flip is related to the real part of the ``mother" GTMD, thus vanishing in the forward limit (it enters the exclusive process scattering amplitudes always multiplied by the transverse momentum transfer $\Delta_\perp$) while the TMD $h_{1}^\perp$ is related to the imaginary part of the same GTMD.       

In a similar way, $\widetilde{H}_T$, involves a double spin flip and its contribution therefore also vanishes in the forward limit. This GPD emerges as the difference between the canonical transversity non-flip amplitudes ($A^{T_Y}$), and the planar transversity non-flip amplitudes ($A^{T_X}$). Since in the forward direction there is no distinction between canonical transversity and planar transversity, clearly then, this GPD requires non-forward scattering to be non-zero. The target polarized transversely at an azimuthal angle different from zero or $\pi/2$ will allow this double correlation to be probed.

Finally, $\widetilde{E}_T$ is the most elusive of the chiral odd GPDs since because of Parity and Time reversal constraints it vanishes for the skewness parameter $\xi=0$, and it is therefore zero in the forward limit; its first moment in $x$ is also zero  \cite{Diehl_odd}.
$\widetilde{E}_T$ can therefore be considered an entirely off-forward product. $\widetilde{E}_T(x, \xi, t)$ describes a transversely polarized quark (along the $x$ axis) in a longitudinally polarized proton. It is T-even and directly connected to the first moment of  the TMD $h_{1L}^\perp$.
Our interest in $\widetilde{E}_T$ stems from the fact that it manifests a similar spin structure to the chiral-even twist three GPD, $G_2$, that was shown to enter the sum rule component for   partonic Orbital Angular Momentum (OAM) in  Ref.\cite{Penttinen,Polyakov}.  Correlated, although indirect information on  OAM 
therefore can be obtained from  both $\widetilde{E}_T$ and $h_{1L}^\perp$ measurements.

In Ref.\cite{GGL_pi0} we evaluated the contributions of the various chiral odd GPDs to the observables in exclusive Deeply Virtual $\pi^0$ electro-Production (DV$\pi^0$P) recently measured at Jefferson Lab \cite{Kub,AvaKim}. In particular,  there exist measurements for the various unpolarized scattering components, for 
 the beam spin asymmetry $A_{LU}$, and for the target spin asymmetries for longitudinal polarization, $A_{UL}$ (unpolarized beam) and $A_{LL}$ (polarized beam). 
%
%
Our analysis uses a flexible parametrization which is based on the reggeized diquark model, that is a spectator model with variable mass of the spectator, $M_X$, which reproduces the Regge behavior in the low $x$, large $M_X$ limit  \cite{GGL,newFF}. The presence of scalar and axial-vector diquarks allows us to model the $u$ and $d$ quark distributions separately, {\it i.e.}  distinguishing between the different isospin projections, $ud$ and $uu$, as it follows from the use of SU(4) symmetry in the nucleon. The model's parameters were determined from quantitative fits of the chiral even GPDs using a compilation of: {\it i)}  flavor separated Dirac and Pauli nucleon form factor data \cite{Cates}, and axial \cite{Schindler} and pseudo-scalar \cite{Fearing} form factor data;  {\it ii)} DVCS data \cite{HallB};  and {\it iii)} reproducing the forward limit of GPDs for both the unpolarized and polarized parton distribution functions from DIS data. 
For the latter we evolved the model from its initial low scale to the scale of the data using leading order Perturbative QCD (PQCD) GPD evolution equations (see {\it e.g.} Refs.\cite{MusRad,GolMar}). 

 For the various asymmetries, in particular, we predicted that the spin modulations which are ideal for measuring chiral odd GPDs using a longitudinally polarized target are 
$A_{UL}^{\sin 2 \phi}$, and the constant in $\phi$ term, $A_{LL}$. The latter are, in fact,  proportional to transverse polarization terms only, and therefore they involve only chiral odd GPDs.  The other modulations, $A_{UL}^{\sin \phi}$, and $A_{LL}^{\cos \phi}$, contain contributions from both longitudinal and transverse photons, their description will always contain a mixture of chiral odd and chiral even GPDs, and thus, they are 
not most appropriate for a clean extraction of the chiral odd sector. Using our model, we found that the dominant contribution to these asymmetries was coming from the chiral even sector (even in the low $Q^2$ kinematical range of Jefferson Lab).         

An additional, important observation is that most of the $A_{UL}$ and $A_{LL}$ data are sensitive to $\widetilde{H}_T$, $E_T$, and $\widetilde{E}_T$, but not directly to $H_T$, and they are therefore not suitable for extracting the tensor charge. 

In this paper we show that the availability of a transversely polarized target and of combined $\eta$ and $\pi^0$ exclusive electroproduction data are both  crucial to extract the tensor charge and its flavor dependence. The extraction from exclusive measurements can in principle allow us to pin down this quantity  more precisely than from SIDIS analyses, an extra advantage being that, as we will explain in what follows,  the low $x$ dependence of transversity which dominates the integration giving $\delta_q$, $(q=u,d)$, will be constrained in exclusive measurements through the $t$ behavior of the Compton Form factors (CFFs).  

Our paper is organized as follows:
in Section \ref{sec:2} we review our approach, and we outline the derivation of the helicity amplitudes entering the cross section 
for deeply virtual pseudoscalar meson production;  in Section \ref{sec:3} we present our results for the various observables and estimate the possibility of extraction of the tensor charge from deeply virtual exclusive experiments;  in Section \ref{sec:conclusions} we draw our conclusions.

\section{Formalism}
\label{sec:2}
We summarize, in  what follows, the formal steps that lead us to parametrize polarized exclusive pseudoscalar meson electroproduction in terms of chiral-odd GPDs. 

\subsection{Definitions and Kinematics} 
We start by defining GPDs at twist-two as the matrix elements of  the following projection of the unintegrated quark-quark proton correlator (see Ref.\cite{Metz1} for a detailed overview),
\footnote{In what follows we can omit the Wilson gauge link without loss of generality \cite{Ji1}.}  
\begin{eqnarray}
W_{\Lambda', \Lambda}^\Gamma(x,\Delta,P) & = & \int \frac{d z^- }{2 \pi} e^{ixP^+ z^-} \left. \langle p', \Lambda' \mid \overline{\psi}\left(-\frac{z}{2}\right) \Gamma \, \psi\left(\frac{z}{2}\right)\mid p, \Lambda \rangle \right|_{z^+=0,{\bf z}_T=0},
\label{matrix}
\end{eqnarray}
where $\Gamma=i\sigma^{i+}\gamma_5 (i=1,2)$
for the chiral odd case, 
and the target's spins are $\Lambda, \Lambda^\prime$. $W_{\Lambda', \Lambda}^\Gamma$ was parametrized as \cite{Diehl_odd},
\begin{eqnarray}
\label{correlator}
W_{\Lambda', \Lambda}^{[i\sigma^{i+}\gamma_5]} & = &  \overline{U}(P',\Lambda') \left[ i \sigma^{+i} H_T(x,\xi,t) +
 \frac{\gamma^+ \Delta^i - \Delta^+ \gamma^i}{2M} E_T(x,\xi,t)   \right. \nonumber \\
& + & \left.  \frac{P^+ \Delta^i - \Delta^+ P^i}{M^2}  \widetilde{H}_T(x,\xi,t)  +
\frac{\gamma^+ P^i - P^+ \gamma^i}{2M} \widetilde{E}_T(x,\xi,t) \right] U(P,\Lambda)    \nonumber \\
& = &   (1-\xi) (\Lambda  \delta_{i1} + i \delta_{i2} ) \delta_{\Lambda,-\Lambda'} H_T + 
  \left( \frac{\Delta_i}{2M} + i \Lambda \, \xi \epsilon^{03ji} \frac{\Delta_j}{2M} \right)  \delta_{\Lambda \Lambda'}  E_T  \nonumber \\ 
& + &  \left[   \frac{\Delta_i}{M} \delta_{\Lambda, \Lambda'}  + (\Lambda \Delta_1+ i  \Delta_2) \frac{\Delta_i}{2M^2} \delta_{\Lambda,- \Lambda'}  \right] \widetilde{H}_T 
+ \left[  \frac{1}{1+\xi} \left( \xi\frac{\Delta_i}{2M} + i \, \Lambda \, \epsilon^{03ji} \frac{\Delta_j}{2M} \right) \delta_{\Lambda \Lambda'}  
+ \xi ( \Lambda \delta_{i1} + i \delta_{i2} ) \delta_{\Lambda,- \Lambda'}  \right ] \widetilde{E}_T \nonumber \\
 \end{eqnarray}
The correlator in Eqs.(\ref{matrix},\ref{correlator}) is expressed in terms of kinematical variables defined in the ``symmetric frame", where we define: $\overline{P}=(p+p')/2$, the average proton momentum, and $\Delta = P-P'$. $\overline{P}$  is along the $z$-axis with momentum, $\overline{P}_3 \approx \overline{P}^+$.  
The four-momenta LC components ($v \equiv (v^+,v^-,\vec{v}_T)$, where $v^\pm=1/\sqrt{2}(v_o \pm v_3)$) are:

\begin{subequations}
\label{kin:sym}
\begin{eqnarray}
\overline{P} & \equiv & \left( \overline{P}^+, \frac{M^2}{\overline{P}^+}, 0) \right) \nonumber  \\
\Delta & \equiv &  \left( \xi \, (2 \overline{P}^+), \frac{ t+ {\bf \Delta}_T^2}{2 \xi \overline{P}^+},  {\bf \Delta}_T ) \right) \\
P & \equiv &   \left((1+\xi) \overline{P}^+,  \frac{M^2+  {\bf \Delta}_T^2/4}{(1+\xi)\overline{P}^+},   {\bf \Delta}_T/2 \right) \nonumber  \\
P' & \equiv &  \left( (1-\xi) \overline{P}^+,  \frac{M^2+  {\bf \Delta}_T^2/4}{(1-\xi)\overline{P}^+},- {\bf \Delta}_T/2  \right), \nonumber  
\label{coord_asym}
\end{eqnarray}
\end{subequations}
where in the DGLAP region (here we consider $x>\xi$) the coordinates of the off-shell struck parton are,
\begin{subequations}
\begin{eqnarray}
k & \equiv &  \left( (x+\xi)\overline{P}^+,  k^-, {\bf k}_T +  {\bf \Delta}_T /2 \right),  \nonumber \\
k' & \equiv &  \left( (x-\xi)\overline{P}^+,  k'^-,{\bf k}_T - {\bf \Delta}_T/2   \right)  
\end{eqnarray}
\end{subequations}
Other useful variables can be written as, 
\[ \hat{s} = (k+q)^2 \approx Q^2(x-\xi)/2\xi , \;\;\;\;  \hat{u} = (k^\prime -q)^2 \approx Q^2 (x+\xi)/2 \xi, \;\;\;\; q^- \approx(Pq)/P^+ = Q^2/(4\xi) (1+\xi)P^+. \]
The loop diagram in Fig.\ref{fig1} integrated over the struck quark's momentum is performed using the variables: $d^4 k \equiv d k^+ d k^- d^2 k_\perp \equiv P^+ dX dk^- d^2 k_\perp$.  
\subsection{Helicity Amplitudes Structure}
To describe spin dependent observables we next introduce the helicity amplitudes
(for a detailed description of the helicity amplitudes formalism in deeply virtual scattering processes see also Ref.\cite{Diehl_hab}). 
For  pseudoscalar meson production one has \cite{AGL,GGL}, 
\begin{eqnarray}
f_{\Lambda_\gamma 0}^{\Lambda \Lambda^\prime} (\xi,t, Q^2)& = & \sum_{\lambda,\lambda^\prime} 
g_{\Lambda_\gamma 0}^{\lambda \lambda^\prime} (x,\xi,t,Q^2)  \otimes
A_{\Lambda^\prime \lambda^\prime, \Lambda \lambda}(x,\xi,t), 
\label{facto}
\end{eqnarray}
where the helicities of the virtual photon and the initial proton are, $\Lambda_\gamma$, $\Lambda$, 
and the helicities of  the produced pion and final proton are $0$, and $\Lambda^\prime$, respectively.
Notice that both longitudinal and transverse polarizations of the virtual photon $\gamma_{L(T)}^*$ can, in principle, contribute. While $\gamma_{L}^*$ was shown to be the leading contribution in Ref.\cite{ColFraStr}, in \cite{AGL,GGL}  a possible scenario beyond collinear factorization was presented for $\gamma_T^* p \rightarrow M p$ which 
explains the large transverse photon polarization contributions  observed in the experimental data in terms of chiral odd GPDs. 
In Eq.(\ref{facto})  we describe the factorization into a ``hard part'', 
$g_{\Lambda_\gamma 0}^{\lambda  \lambda^\prime}$ for the partonic subprocess 
$\gamma^*_T + q \rightarrow \pi^0 + q$, which appears now at twist three,
and  the quark-proton helicity amplitudes, $A_{\Lambda^\prime,\lambda^\prime;\Lambda,\lambda}$ 
that contain the chiral odd GPDs.

The amplitudes $A_{\Lambda^\prime \lambda^\prime, \Lambda \lambda}$ implicitly involve an integration over the unobserved quark's transerve momentum, $k_T$,
and are functions of 
$x_{Bj} =Q^2/2M\nu \approx 2\xi/(1-\xi^2)$, $t$ and $Q^2$. The convolution integral in  Eq.(\ref{facto}) 
 is given by $\otimes \rightarrow \int_{-1}^1 d x$.

The connection with the correlator is carried out by considering,
\begin{eqnarray}
A_{\Lambda' \lambda', \Lambda \lambda} = \int \frac{d z^-}{2 \pi} e^{ixP^+ z^-} \left. \langle p', \Lambda' \mid {\cal O}_{\lambda' \lambda}(z) \mid p, \Lambda \rangle \right|_{z^+=0, {\bf z}_T=0}, 
\end{eqnarray} 
where,
\begin{eqnarray}
{\cal O}_{-+}(z) & = & -i \bar{\psi}\left(-\frac{z}{2}\right) (\sigma^{+1} - i \sigma^{+2})  \psi\left(\frac{z}{2}\right) \\
{\cal O}_{+ -}(z) & = & i \bar{\psi}\left(-\frac{z}{2}\right) (\sigma^{+1} + i \sigma^{+2}) \psi\left(\frac{z}{2}\right), 
\end{eqnarray}
By taking this into account in Eq.(\ref{correlator}), and by adding and subtracting the expressions corresponding to $i=1,2$, respectively, one obtains the expressions for  the chiral odd helicity amplitudes in terms of 
GPDs   \cite{Diehl_odd,Diehl_hab},
\begin{subequations}
\label{GPDodd}
\begin{eqnarray}
A_{++,--} & = &  \sqrt{1-\xi^2}  \left[ { H}_ T + \frac{t_0-t}{4M^2} \widetilde{ H}_T      
+ \frac{\xi^2}{1-\xi^2}  { E}_T  + \frac{\xi}{1-\xi^2} \widetilde{ E}_T \right]   \\ 
A_{+-,-+} & = &  -  \sqrt{1-\xi^2}  \,  \frac{t_0-t}{4M^2} \, \widetilde{ H}_T \\
A_{++,+-} & = & \frac{\sqrt{t_0-t}}{4M} \left[ 2\widetilde{ H}_T  + (1-\xi)  \left({ E}_T - \widetilde{ E}_T \right) \right]  \\
A_{-+,--} & = & \frac{\sqrt{t_0-t}}{4M}  \,  \left[  2\widetilde{ H}_ T + (1+\xi) \left( { E}_T + \widetilde{ E}_T \right) \right].
\end{eqnarray}
\end{subequations}
Notice that $A_{+-,++}$, $A_{++,+-}$ change sign under Parity while $A_{--,++} $, $A_{+-,-+} $, do not change sign; since $g_{10}^{+-}$ also changes sign, then $f_{10}^{++}$,  $f_{10}^{--}$ will not change sign under Parity, while   $f_{10}^{+-} $, and$ f_{10}^{+-} $ will change sign.

The chiral-odd coupling at the pion vertex for the subprocess $\gamma^* q \rightarrow \pi^0 q'$  
 is given by,
\begin{eqnarray} 
g_{\Lambda_\gamma 0}^{\lambda  \lambda^\prime} & =  & g_\pi^{V(A)}(Q^2)   \,  q^- 
\left[ \bar{u}(k^\prime,\lambda^\prime) \gamma^\mu \gamma^+  \gamma_5 u(k,\lambda) \right]  
 \epsilon_\mu^{\Lambda_\gamma}   \left( \frac{1}{\hat{s} - i \epsilon } - \frac{1}{\hat{u} - i \epsilon} \right).
\label{g_odd}
\end{eqnarray}
where we distinguish three different contributions: from the term, 
$K = q^-[1/(\hat{s}-i  \epsilon) -1/(\hat{u} - i \epsilon)]$,
\begin{eqnarray}
K = \frac{Q^2}{2 x_{Bj} P^+} \, \frac{x_{Bj}}{Q^2} \, C^+ \equiv \frac{1}{2P^+} C^+ \\
C^+ =  \frac{1}{x- \xi + i \epsilon }  + \frac{1}{x+\xi - i \epsilon };
\end{eqnarray}
from the contraction,
\begin{eqnarray}
\bar{u}(k^\prime,\lambda^\prime) \gamma^\mu \gamma^+  \gamma_5 u(k,\lambda)    \epsilon_\mu^{\Lambda_\gamma}   & = & 
 N N^\prime \,  {\rm Tr} 
\left\{ (\!\not{k}  +m) \, \hat{\mathcal O}_{\lambda,\lambda^\prime} (\!\not{k}^\prime + m) \gamma^\mu \gamma_5 \gamma^+  \right\}  \epsilon_\mu^{\Lambda_\gamma}  \nonumber \\
& \approx &  -\frac{1}{\sqrt{k^{\prime \, +} k^+}} \, \left[ k^o p^{\prime \, +} -(k k^\prime) + k^+ k^{\prime \, o} \right] (\epsilon_1^{+1} - i \epsilon_2^{+1})  = \sqrt{(x-\xi)(x+\xi)} P^+
\end{eqnarray}
where $N = 1/\sqrt{P^+(x+\xi)}$ and $N^\prime=1/\sqrt{P^+(x-\xi)}$ are the quark spinors normalizations (details are given in Appendix \ref{appa}), and,
\begin{subequations}
\begin{eqnarray}
\hat{\mathcal O}_{\pm \pm } = \frac{1}{4}(1+\gamma^o) (1\pm \gamma_5 \gamma_3) && \\ 
 \hat{\mathcal O}_{\pm \mp}  = -\frac{1}{4}(1+\gamma^o) \gamma_5(\gamma_1 \mp i\gamma_2), && 
 \end{eqnarray}
\end{subequations}
and finally 
from the $Q^2$ dependent  form factor $g_\pi^{V(A)}(Q^2)$ where we separate \cite{AGL}
the $J^{PC}=1^{- -}$ (V) and $J^{PC}=1^{+ -}$ (A), $t$-channel exchanges in the  amplitudes for transverse and longitudinal virtual photons, respectively. 
The two distinct contributions arise when one goes beyond a simple one gluon exchange description of the chiral odd coupling $\propto \gamma^5$.    
\footnote{These two distinct configurations are the dominant terms in the  two  series with  {\it natural parity} one ($1^{--}, 3^{--} ... $), labeled $V$, and 
{\it unnatural parity} one ($1^{+-}, 3^{+-} ...$), labeled $A$. }
What makes the two contributions $\gamma^* (q \bar{q})_V \rightarrow \pi^0$ and 
$\gamma^* (q \bar{q})_A \rightarrow \pi^0$ distinct  
is that, in the natural parity case (V), L is always the same for the initial and final states, or $\Delta L=0$,
while for unnatural parity (A), $\Delta L =1$. 
We modeled this difference by replacing the collinear factorization expressions  with the following expressions containing a modified kernel 
\begin{eqnarray}
g^V_{\Lambda_{\gamma^*},\lambda; 0, \lambda^\prime}  =  \int dx_1 dy_1 \int  d^2 b  
\, \hat{\psi}_V(y_1,b) \, \hat{{\cal F}}_{\Lambda_{\gamma^*},\lambda; 0, \lambda^\prime}(Q^2,x_1,x_2,b) \alpha_S(\mu_R)
 \exp[-S]     \, \hat{\phi}_{\pi^0}(x_1,b)  && \\
g^A_{\Lambda_{\gamma^*},\lambda; 0, \lambda^\prime}  =  \int dx_1 dy_1 \int d^2  b  
\, \hat{\psi}_A(y_1,b) \, \hat{{\cal F}}_{\Lambda_{\gamma^*},\lambda; 0, \lambda^\prime}(Q^2,x_1,x_2,b) \alpha_S(\mu_R)
\exp[-S]  \, \hat{\phi}_{\pi^0}(x_1,b) &&
\end{eqnarray}
where, 
\begin{equation}
\hat{\psi}_{A}(y_1,b) = \int d^2 k_T J_1(y_1 b) \psi_V(y_1,k_T) 
\end{equation}
The higher order Bessel function describes the situation where $L$ is always larger in the initial state. 
In impact parameter space this corresponds to configurations of larger radius. 
By considering the dominant LC components, we see that only 
$g_{10}^{+-}$ survives, 
\begin{eqnarray}
\label{ampg2}
g_{10}^{+-}  & \approx & g_\pi^{V(A)}(Q^2)   \: C^+
\end{eqnarray}


\begin{figure}
\includegraphics[width=9.cm]{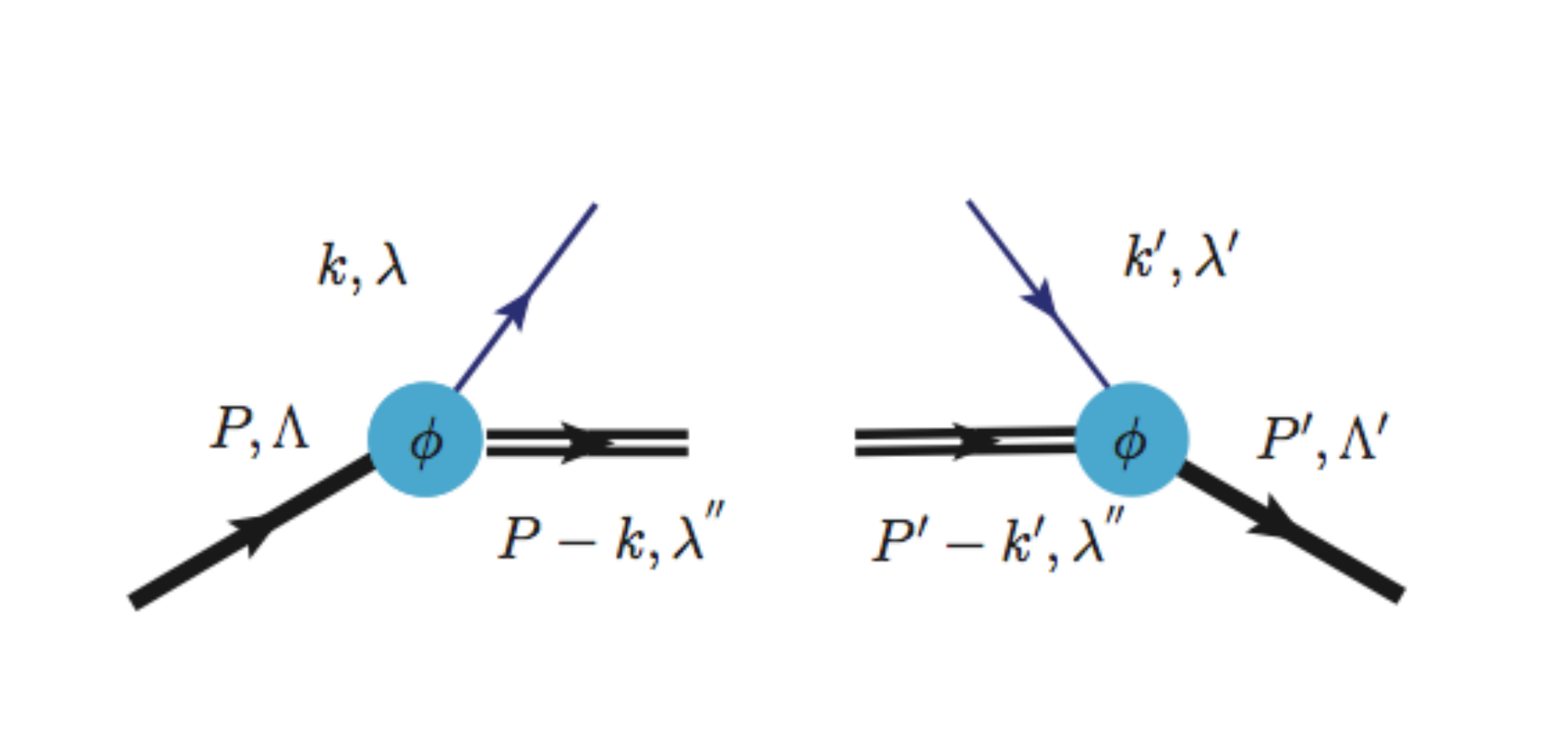}
\caption{Vertex structures defining the spectator model tree level diagrams.  
}
\label{fig_LCWF}
\end{figure}

Putting together all steps,  we can write a calculable form for the convolution in Eq.(\ref{facto}).
For a transverse photon we obtain,
\begin{subequations}
\label{helamps_gpd2}
\begin{eqnarray}
f_{10}^{++} &= &  g_\pi^{V}(Q) \frac{\sqrt{t_0-t}}{4M}  \,  \left[  2\widetilde{\cal H}_ T + (1+\xi) \left( { \cal E}_T + \widetilde{\cal E}_T \right) \right]  
 \\
f_{10}^{+-} & = &   \frac{g_\pi^{V}(Q)+ g_\pi^{A}(Q)}{2} \, \sqrt{1-\xi^2}  \left[ { \cal H}_ T + \frac{t_0-t}{4M^2} \widetilde{ \cal H}_T      
+ \frac{\xi^2}{1-\xi^2}  {\cal  E}_T  + \frac{\xi}{1-\xi^2} \widetilde{\cal  E}_T \right]   \\
 f_{10}^{-+} & = &  -   \frac{g_\pi^{A}(Q)- g_\pi^{V}(Q)}{2}  \, \sqrt{1-\xi^2}  \,  \frac{t_0-t}{4M^2} \, \widetilde{\cal  H}_T 
 \\
f_{10}^{--} & = & g_\pi^{V}(Q)   \frac{\sqrt{t_0-t}}{4M} \left[ 2\widetilde{ \cal H}_T  + (1-\xi)  \left({\cal  E}_T - \widetilde{\cal  E}_T \right) \right] 
\end{eqnarray} 
\end{subequations}
where the matching of the $V$ and $A$ contributions to the helicity amplitudes is as follows: $f_{10}^{++}, f_{10}^{--} \propto g^V$, $f_{10}^{+-}  \propto g^V+g^A$, 
$f_{10}^{-+} \propto g^V-g^A$ (see Ref. \cite{GGL_pi0short,GGL_pi0}).
 ${\cal H}_T$, etc., are the convolutions of the GPDs with $C^+$, or the Compton form factors which at leading order in PQCD are given by,  
\begin{equation}
\label{CFF_def}
{\cal F}_T(\xi,t,Q^2) = \int_{-1}^1 dx \; C^+ \, F_T(x,\xi ,t,Q^2) 
\end{equation}
$F_T \equiv {\cal H}_T, {\cal E}_T, \widetilde{\cal H}_ T, \widetilde{\cal E}_ T$.

\subsection{Chiral Odd GPDs from Helicity Amplitudes}
\begin{figure}
\includegraphics[width=6.cm]{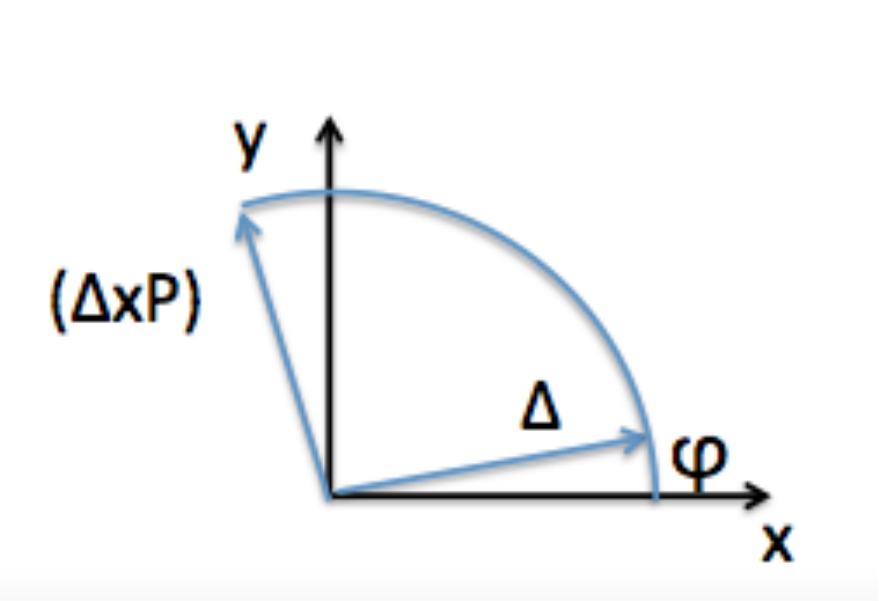}
\caption{$\Delta$ and $({\bf \Delta} \times {\bf P})/\mid {\bf P} \mid$ which defines transversity.  In this paper we take $\Delta$ along the $x$-axis without loss of generality. 
}
\label{fig3}
\end{figure}

The chiral odd CFFs (or GPDs) are obtained by inverting Eqs.(\ref{helamps_gpd2}) (or Eqs.(\ref{GPDodd})). 
For instance, for $\xi=0$ one has, 
\footnote{Numerical calculations throughout this paper were conducted using the full $\xi$ dependent expressions, the expressions above are shown for simplicity.}
\begin{subequations}
\label{gpdsodd:eq}
\begin{eqnarray}
\label{AHTbar}  
\frac{ \sqrt{t_o-t}}{2M}   \left[ 2 \widetilde{H}_T(x, 0, t)  + E_T(x,0,t) \right]  & = &  A_{++,+-}   + A_{-+,--} \nonumber \\
& = &  A^{T_Y}_{++,++} - A^{T_Y}_{+-,+-} + A^{T_Y}_{-+,-+} - A^{T_Y}_{--,--}    \\
 \label{AHTX} 
  H_T(x, 0, t) & = & A_{++,--}  + A_{-+,+-} \nonumber \\ 
& = &   A^{T_Y}_{++,--} - A^{T_Y}_{+-,-+} + A^{T_Y}_{--,++} - A^{T_Y}_{-+,+-} \nonumber \\
& = &   A^{T_X}_{++,++} - A^{T_X}_{+-,+-} - A^{T_X}_{-+,-+} + A^{T_X}_{--,--}   \\
\label{AHTtildeX}
-\frac{t_o-t}{4M^2}  \widetilde{H}_T(x, 0, t)   & = &  A_{-+,+-} \nonumber \\  
& = &  A^{T_Y}_{++,++} - A^{T_Y}_{+-,+-} - A^{T_Y}_{--,--} + A^{T_Y}_{-+,-+}  +  A^{T_Y}_{++,--} - A^{T_Y}_{+-,-+} + A^{T_Y}_{--,++} - A^{T_Y}_{-+,+-} \nonumber  \\
& = &  A^{T_X}_{++,++}+A^{T_X}_{+-,+-}+A^{T_X}_{-+,-+}+A^{T_X}_{--,--}   \\
- \frac{ \sqrt{t_o-t}}{2M}  \widetilde{E}_T(x, 0, t) & = &  A_{++,+-}   - A_{-+,--}   \nonumber \\
& = &   A^{L,T_X}_{++,++}-A^{L,T_X}_{+-,+-}-A^{L,T_X}_{-+,-+}+A^{L,T_X}_{--,--}  =0 
\end{eqnarray}
\end{subequations}
where we show the GPDs calculated using the helicity amplitudes (first line of each equation), and using the transversity bases, $T_Y$, with the transverse spin orthogonal to ${\bf \Delta}$ (without loss of generality ${\bf \Delta}$ is assumed to be along the $x$ axis), the planar-transversity basis $T_X$, with transverse spin along $x$, and a mixed longitudinal and planar-transverse basis, $L,T_X$ with longitudinal an transverse along $x$ spins in the initial and final states, respectively.
If the scattering plane for the quark+nucleon is the $x-z-$plane (with ${\bf P}$ and ${\bf \Delta}$) and the $y-$direction is along ${\bf P} \times {\bf \Delta}$, then for each particle's spin $\frac{1}{2}$ helicity state, $\sqrt{2} \mid T_X=\pm \rangle = \mid + \frac{1}{2} \rangle \pm \mid - \frac{1}{2} \rangle $ and $\sqrt{2} \mid T_Y=\pm \rangle = \mid + \frac{1}{2} \rangle \pm i \mid - \frac{1}{2} \rangle $ (Fig.\ref{fig3}). 
Note that $\widetilde{E}_T(x, \xi, t)$ vanishes for $\xi \rightarrow 0$ because the two helicity amplitudes become equal due to parity and time reversal invariance. 

In order to give a partonic interpretation, the spin structures on the RHS and LHS of Fig.\ref{fig1}  need to be the same, or diagonal in spin, where the direction of transverse spin is established using the bases defined with $\Delta$. For instance, $H_T$ is diagonal in $T_X$, while $\widetilde{H}_T$ is diagonal in a $T_X$ and $T_Y$ mixed basis. 

\subsubsection{$\overline{E}_T = 2 \widetilde{H}_T  + E_T$}
By inspecting the spin content of Eqs.(\ref{gpdsodd:eq}) we see that Eq.(\ref{AHTbar}) corresponds to the same combination as for  the Boer Mulders function $h_1^\perp$ \cite{BoerMul}.  
This is well known to be vanishing at leading order, in the absence of a gauge link, owing to the ``naive" T-odd nature of $h_1^\perp$ (see \cite{BarDraRat} for a review).  The question of whether $h_1^\perp$ and $\left[ 2 \widetilde{H}_T(x, 0, t)  + E_T(x,0,t) \right]$ could be related was initially posed in Ref.\cite{Bur1} within a transverse coordinate space representation.
A more general framework from which to address this question was subsequently provided in Ref.\cite{Metz1} using the GTMDs, or unintegrated over ${\bf k}_T$ TMDs.  GTMDs  are complex objects that can be parametrized as
\begin{equation}
X(x,k_T,\xi,t;Q^2) = X^e + i X^o
\label{GTMD}
\end{equation}
where $X$ represents any twist two GPD, and $X^e$, and $X^o$, the real and imaginary parts, are also the T-even and T-odd components, respectively. The leading twist GPDs are obtained by integrating Eq.(\ref{GTMD}); they are  T-even and they  can derive only from the real parts of the amplitude combinations. On the other side, the leading twist TMDs, are obtained in the forward limit ($t,\xi \rightarrow 0$), and they can also be T-odd (within the well known appropriate interpretation of the gauge links)
\footnote{A discussion of gauge links is beyond the scope of this paper and is omitted here}
 in which case they involve the imaginary part of the amplitudes combinations. In this context we see that  both $h_1^\perp$ and  $\overline{E}_T =  2 \widetilde{H}_T(x, \xi, t;Q^2)  + E_T(x,\xi,t;Q^2) $ participate in the same equation at the GTMD level, 
\begin{equation}
\overline{E}_T(x,k_T,\xi,t;Q^2)   = \Re e  \overline{E}_T  (x,k_T,\xi,t;Q^2) + i \, \Im m \overline{E}_T (x,k_T,\xi,t;Q^2) 
\label{GTMD2}
\end{equation}
with,
\begin{eqnarray}
 h_1^\perp(x,k_T) \equiv \underset{t,\xi \rightarrow 0}{\rm lim}  \frac{\sqrt{t_0-t}}{2M} \, \overline{E}_T (x,k_T,\xi,t;Q^2) & = &  \, \Im m \overline{E}_T (x,k_T,0,0;Q^2)    \\
\overline{E}_T (x,\xi,t;Q^2)  \equiv  \int^{1}_{-1} d^2 k_T \overline{E}_T (x,k_T,\xi,t;Q^2)& = &  \int^{1}_{-1} d^2 k_T \Re e  \overline{E}_T (x,k_T,\xi,t;Q^2). 
\label{limitsGTMD}
\end{eqnarray}
(our formalism differs from Ref.\cite{Metz1} where the GTMD $\overline{E}$ is further separated into terms arising from different Lorentz structures while for simplicity, yet  keeping general, we use only one term, $\overline{E}_T(x,k_T,\xi,t;Q^2)$). 

We conclude that $h_1^\perp(x,k_T)$ and $\overline{E}_T (x,\xi,t;Q^2)$, although they derive from the same helicity amplitudes combinations, will differ, or they are unrelated, in general. In other words, $h_1^\perp(x,k_T)$ and $\overline{E}_T (x,\xi,t;Q^2)$ provide separate information from the imaginary and real parts of the amplitudes, respectively. $\pi^0$ and $\eta$ exclusive electroproduction data allow for an independent extraction of $\overline{E}_T$, and they are, therefore,  ideal for testing this aspect of the theory.   

Integrating Eq.(\ref{limitsGTMD}) over $x$ at $t=0$, one obtains  the tensor anomalous magnetic moment \cite{Bur1},
\begin{eqnarray}
\kappa^T_q = \int_{-1}^1 dx \overline{E}_T(x,0,0;Q^2) 
\label{kappaT}
\end{eqnarray}
This can also be extracted from the analysis of exclusive $\pi^0$ and $\eta$ electroproduction data as we explain below.

\subsubsection{Transversity}
By inspecting the spin structure of Eqs.(\ref{AHTX},\ref{AHTtildeX}) we see that $H_T$ is diagonal in planar transversity, whereas ${\widetilde H}_T$ 
is neither diagonal in canonical transversity nor planar transversity.  However, 
subtracting Eq.(\ref{AHTtildeX}) from Eq.(\ref{AHTX}) one obtains a diagonal combination of $T_Y$ amplitudes, 
\begin{eqnarray}
H^\prime_T & = & H_T + \frac{t_0 - t}{2M^2} {\widetilde H}_T 
 =  A_{++,++}^{T_Y} + A_{--,--}^{T_Y} - A_{+-,+-}^{T_Y} - A_{-+,-+} ^{T_Y}. 
 \label{HTprime2}
 \end{eqnarray}
This is the analog of  the TMD relation, $h_1(x,{\bf k}_T^2)=h_{1T}(x,{\bf k}_T^2) + ({\bf k}_T^2 / 2M^2) h_{1T}^\perp({\bf k}_T^2)$ where, owing to the fact that
$\Delta =0$, all functions that were diagonal in a given direction become diagonal on the transverse plane for any direction ({\it i.e.} $\Delta$ is no longer there to define the direction). 
Therefore, in the forward direction, $H_T^\prime = H_T$, and this is diagonal in either $T_X$ or $T_Y$, consistently with the definition of transversity,    
\begin{eqnarray}
H_T(x,0,0; Q^2) & = & h_1(x,Q^2) 
\end{eqnarray}     
Integrating over $x$ one obtains  the tensor charge, at $t=0$,
\begin{eqnarray}
\delta_q = \int_{-1}^1 dx H_T(x,0,0;Q^2) 
\end{eqnarray}
$\widetilde{H}_T$ can also be interpreted in a mixed planar/transversity basis as the distribution of transversely polarized quarks along $x$ in a transversely polarized proton along $y$.  
In fact, Eq.(\ref{AHTtildeX})  is related to  the first $k_T$ moment of $h_{1T}^\perp$,
\begin{eqnarray}
\underset{t \rightarrow 0}{\rm lim} \, \widetilde{H}_T(x,0,0, Q^2)  & \rightarrow & h_{1T}^{\perp (1)}(x,Q^2). 
\end{eqnarray}
It is important to keep in mind that although this relation holds (modulo an $x$ dependent factor) when tested using simple spectator models \cite{Metz2}, in Ref. \cite{Metz1} it has been disproven based on the GTMD substructures underlying $\widetilde{H}_T(x,0,0, Q^2)$, and  $h_{1T}^{\perp (1)}(x,Q^2)$, which differ from one another. The physical motivation of such a discrepancy obtained in parametrizing the chiral odd GTMDs correlator is however unclear to date. Whether this is an artifact of the proposed parametrization or it can be traced back to different spin configurations,  is a subject for further exploration.

\subsubsection{$\widetilde{E}_T$}
$\widetilde{E}_T(x, \xi, t)$ describes a transversely polarized quark (along the $x$ axis) in a longitudinally polarized proton. It is T-even and directly connected to a TMD, being related to the first moment of  $h_{1L}^\perp$,
\begin{equation}
\int d^2 k_T h_{1L}^\perp(x,k_T) \propto  \widetilde{E}_T(x, 0, t)\mid_{t \rightarrow 0} = 0
\end{equation}
Although $\widetilde{E}_T$ is 0  for $\xi=0$ due to Parity and Time reversal constraints,
$\widetilde{E}_T(x, \xi, t)$ can, however,  be measured (see also our paper on $\pi^0$ electroproduction, Ref.\cite{GGL_pi0}). 
What makes $\widetilde{E}_T$ interesting is that its spin structure is similar to the one that appears in the chiral-even twist three GPD, $G_2$, that was shown to enter the sum rule component for   partonic Orbital Angular Momentum (OAM) in  Ref.\cite{Polyakov}.  Several candidates among the TMDs and GPDs were proposed recently \cite{Ma,Schmidt,LorPas} as observables for the OAM component.  In Ref. \cite{OAM} we illustrated the helicity amplitude structure of partonic OAM, and we confirmed from an alternative perspective, that OAM is twist three, and that it is uniquely observable through DVCS type measurements of  $G_2$ (see also \cite{Hatta}). In particular, we pointed out that    
based on Parity constraints, the twist two chiral even GTMD labeled $F_{14}$ in \cite{Metz1}, describing an unpolarized quark in a longitudinally  polarized proton, could  not contribute  to OAM. We also pointed out that another possible candidate, the chiral odd TMD, $h_{1T}^\perp$ \cite{Efremov}, does not display the necessary spin structure of partonic OAM.  To confirm this picture, it would be important to obtain correlated, although indirect information on  OAM  from  both $\widetilde{E}_T$ and $h_{1L}^\perp$ measurements.

\subsection{Chiral Odd GPDs in Spectator Model}
\label{sec:gpdsodd}
Eqs.(\ref{helamps_gpd2}) provide the helicity amplitudes that enter directly the observables for pseudoscalar meson electroproduction. 
However, in practical calculations we find an outstanding problem:  differently from the chiral even sector, even using available models for the chiral odd GPDs both the normalization to the form factors and the forward limit for these quantities are largely undetermined from independent measurements (with the exclusion, perhaps, of transversity, $h_1$). 
This makes it difficult to estimate the magnitude of the chiral odd GPDs. In the spectator model, that we describe below, one can overcome this difficulty by exploiting Parity relations among the helicity amplitudes that allow us to connect the chiral odd GPDs to the chiral even ones (analogous relations were found to hold for TMDs in Ref.\cite{BCR}). Although this is a model dependent procedure, we consider it a necessary step at an initial stage of both theoretical and experimental investigations. 

To extract the chiral odd GPDs from DV$\pi^0(\eta)$P data we propose a parametrization based on the  reggeized diquark  model developed in Refs.\cite{AHLT1,AHLT2,GGL,newFF}. This is, in a nutshell, 
a spectator model with variable mass of the spectator, $M_X$, which allows us to reproduce the Regge behavior in the limit $M_X \rightarrow \infty$  ($x \rightarrow 0$). 
The model was evolved from its initial low scale to the scale of the data using leading order Perturbative QCD (PQCD) GPD evolution equations (see {\it e.g.} Refs.\cite{MusRad,GolMar}). 
In the chiral even sector we determined the model's parameters from a fit using a compilation of three different data sets, namely, nucleon form factor data (flavor separated Dirac and Pauli form factors \cite{Cates}, axial \cite{Schindler} and pseudo-scalar form factor \cite{Fearing}), DVCS data \cite{HallB},  and DIS data on $F_2^p(n)$. 
For the latter  we accurately reproduced the valence quarks PDFs obtained from global fits \cite{PDFS}.  
In Ref.\cite{GGL} our approach lead us to succesfully reproduce data on various observables in DVCS besides the ones used in the fit, namely the charge \cite{HERMES1,HERMES2},  and transverse \cite{HERMES1,HERMES2} single spin asymmetries. 

In the chiral odd sector the GPDs are largely unconstrained. This is mostly due to the fact that, differently from the GPDs in the chiral even sector, 
 no experimental measurements from $t$ dependent form factors exist, which would provide normalizations for the GPDs. In the analysis of $\pi^0$ and $\eta$ electroproduction data it is, however, important to be able to gauge the size of the various chiral odd GPDs contributions.

We therefore developed an alternative procedure by exploiting Parity relations within the reggeized diquark model.
The extension of this model to the chiral odd sector is explained in detail in Ref.\cite{GGL_pi0} where 
we obtained predictions for both the unpolarized and longitudinally polarized observables in $\pi^0$ electroproduction, namely the beam spin asymmetry, $A_{LU}$, and the polarized target asyemmetries, $A_{UL}$ and $A_{LL}$. 

The Parity relations between chiral even and chiral odd GPDs which are valid, in general, within a class of models including any type of spectator model with diquark spin $S=0,1$ and angular momentum $L=0$
\footnote{More complicated intermediate state configurations with $L\neq 0$ have been considered recently in Ref.\cite{Pena}. The Parity relations would be different in this case.}
  yield, respectively, for $S=0$,
\begin{subequations}
\label{Amp0}
\begin{eqnarray}
A_{++,--}^{(0)}   & =  & A_{++,++}^{(0)}   \\
A_{++,+-} ^{(0)}    &  = &  - A_{++,-+}^{(0)*}  \\ 
A_{+-,++}^{(0)}   &  = &  - A_{-+,++}^{(0)*}  \\
\label{f3}
A_{+-,-+}^{(0)}  &= &\frac{t_0-t}{4M} \sqrt{1-\xi^2} \frac{\tilde{X}}{m+MX^\prime}  \left[ E - \frac{\xi}{1-\xi^2}  \widetilde{E} \right],
\end{eqnarray}
\end{subequations}
and, for $S=1$,  
\begin{subequations}
\label{Amp1}
\begin{eqnarray}
A_{++,--}^{(1)}  & = &   \displaystyle\frac{X+X^\prime}{1+ X X^\prime} \; A_{++,++} ^{(1)} \\
A_{+-,-+}^{(1)}  & = & 0 \\
A_{++,+-}^{(1)}  & = & - \sqrt{ \frac{\langle \tilde{k}_\perp^2\rangle /P^{+ \, 2} }{X^{\prime \, 2} + \langle \tilde{k}_\perp^2 \rangle /P^{+ \, 2} }  } \; A_{++,-+} ^{(1)*} \\
A_{+-,++}^{(1)}  & = & - \sqrt{ \frac{\langle k_\perp^2  \rangle /P^{+ \, 2} }{X^2 + \langle k_\perp^2 \rangle/P^{+ \, 2} }  } \; A_{-+,++}^{(1)*},
\end{eqnarray}
\end{subequations}
The relations in Eqs.(\ref{Amp0}) are valid only if one of the two $\phi$ functions is real. By using Parity symmetry one cannot connect directly the chiral odd amplitude $A_{+-,-+}$, with its chiral even counterpart $A_{+-,+-}$ since both involve complex $\phi$ functions. Physically this corresponds to the fact that $A_{+-,-+}$
involves a double spin flip, and it must therefore be proportional to $\Delta_\perp^2 $, while $A_{+-,+-}$ is non-flip. 
$A_{+-,-+}$ is, therefore, evaluated directly in Eq.(\ref{f3}). 
Eqs.(\ref{Amp1}) were obtained by taking into account the angular momentum exchange between the LHS and RHS vertices in Fig.\ref{fig1} so that each amplitude on the LHS (chiral odd) can no longer be obtained as a simple product of the two vertices which give the chiral even amplitude on the RHS. 

We can then obtain the chiral odd GPDs sets $F_T^{(0),(1)}$, by 
inverting the expressions of the quark parton helicity amplitudes in both the chiral even \cite{Diehl_hab}, and chiral odd (\cite{Diehl_hab} and Eqs.(\ref{GPDodd})) sectors.  

In Ref.\cite{GGL_pi0} we obtained, for $S=0$,
\begin{subequations}
\label{S0}
\begin{eqnarray}
\widetilde{H}_T^{(0)} & = &   -\frac{1}{F} \left(E^{(0)}  -  \frac{\zeta}{2} \widetilde{E}^{(0)}  \right) \\ 
E_T^{(0)} & = &  \frac{(1-\zeta/2)^2}{1-\zeta} \left[E^{(0)} - 2 \widetilde{H}_T^{(0)} - (\frac{\zeta/2}{1-\zeta/2})^2 \widetilde{E}^{(0)}  \right]  \\
\widetilde{E}_T^{(0)} & = &  \frac{\zeta/2 (1-\zeta/2)}{1-\zeta} \left[E^{(0)}  - 2\widetilde{H}_T^{(0)}   - \widetilde{E}^{(0)}  \right] \\
H_T^{(0)}  & = &  \frac{H^{(0)}  + \widetilde{H}^{(0)} }{2} - \frac{\zeta^2/4}{1-\zeta}\frac{E^{(0)} + \widetilde{E}^{(0)} }{2} -  \frac{\zeta^2/4}{(1-\zeta/2)(1-\zeta)} E_T^{(0)}  + 
\frac{\zeta/4 (1-\zeta/2)}{1-\zeta} \widetilde{E}^{(0)} _T -
\frac{t_0-t}{4M^2} \frac{1}{F} \left(E^{(0)}  -  \frac{\zeta}{2} \widetilde{E}^{(0)} \right) \nonumber \\
\end{eqnarray}
\end{subequations}
and for $S=1$, 
\begin{widetext}
\begin{subequations}
\label{S1}
\begin{eqnarray}
\widetilde{H}_T^{(1)} & = &  0   \\
E_T^{(1)} & = &    \frac{1-\zeta/2}{1-\zeta} \left[ \tilde{a} \left(E^{(1)} -  \frac{\zeta/2}{1-\zeta/2} \widetilde{E}^{(1)} \right)   + a \left(E^{(1)} + \frac{\zeta/2}{1-\zeta/2} \widetilde{E}^{(1)} \right) \right] \\
\widetilde{E}_T^1 & = &   \frac{1-\zeta/2}{1-\zeta} \left[ \tilde{a} \left(E^{(1)} -  \frac{\zeta/2}{1-\zeta/2} \widetilde{E}^{(1)} \right)  -  a \left(E{(1)} + \frac{\zeta/2}{1-\zeta/2} \widetilde{E}^{(1)} \right) \right]  \\
H_T^{(1)} & = &  - G \left[  \frac{H^{(1)} + \widetilde{H}^{(1)}}{2} - \frac{\zeta^2/4}{1-\zeta}\frac{E^{(1)} + \widetilde{E}^{(1)}}{2} \right] -  \frac{\zeta^2/4}{1-\zeta} E_T^{(1)} + 
\frac{\zeta/4}{1-\zeta} \widetilde{E}_T^{(1)}  
\end{eqnarray}
\end{subequations}
\end{widetext}
where we used the following variables: $\zeta = 2 \xi/(1+\xi)$, $X=(x+\xi)/(1+\xi)$, $X^\prime = X-\zeta$, and the various kinematical factors are, 
\[ F= \left(M X^\prime  + m_q \right) \, \frac{1-\zeta}{\widetilde{X} },  \;\;\;\;    G= \displaystyle\frac{X+X^\prime}{1+ X X^\prime}, \;\;\;\; \widetilde{X} = \frac{1-X}{1-\zeta} \]
and
\[a =  \displaystyle\sqrt{ \frac{\langle k_\perp^2 \rangle}{X^2 + \langle k_\perp^2 \rangle \left(\frac{2M\zeta^2}{Q^2}\right)^2} }, \;\;\;\;
\tilde{a} =   \displaystyle\sqrt{ \frac{\langle \tilde{k}_\perp^2 \rangle}{(X-\zeta)^2 + \langle \tilde{k}_\perp^2 \rangle \left(\frac{2M\zeta^2}{Q^2}\right)^2} } \]
where $\langle k_\perp^2 \rangle^{1/2} \approx 0.3$ GeV, and  $\langle \tilde{k}_\perp^2 \rangle^{1/2} = \langle ({\bf k}_\perp + (1-X) {\bf \Delta})^2 \rangle^{1/2} \approx \langle k_\perp^2 \rangle^{1/2} + (1-X) \mid {\Delta}\mid $.

The chiral even GPDs used to evaluate Eqs.(\ref{S0}) and (\ref{S1}) were taken from the parametrization developed in Ref.\cite{newFF}.  In the forward limit one obtains,
 \begin{subequations}
\label{S00}
\begin{eqnarray}
\widetilde{H}_T^{(0)} & = &   - \frac{M(1-x)}{m+Mx} E^{(0)}  \\ 
E_T^{(0)} & = &   \frac{M(2-x) + m}{m+Mx}  E^{(0)}   \\
\widetilde{E}_T^{(0)} & = &  0 \\
H_T^{(0)}  & = &  \frac{H^{(0)}  + \widetilde{H}^{(0)} }{2} 
\end{eqnarray}
\end{subequations}
and for $S=1$, 
\begin{subequations}
\label{S10}
\begin{eqnarray}
\widetilde{H}_T^{(1)} & = &  0   \\
E_T^{(1)} & = &   2 a E^{(1)}  \\
\widetilde{E}_T^1 & = &   0 \\
H_T^{(1)} & = & -  \frac{2x}{1+x^2}  \frac{H^{(1)} + \widetilde{H}^{(1)}}{2} 
\end{eqnarray}
\end{subequations}
Note that once the chiral odd GPDs are calculated using the equations above, at a given initial scale of the model, $Q_o^2 \lesssim 1$ GeV$^2$, 
their evolution to the scale of the data must proceed according to the equations developed for the chiral odd sector in \cite{odd_evol,Kumano}. 
The parametric form for the chiral even GPDs was given in Ref.\cite{newFF }. 
As both new DVCS and meson electroproduction data become available, it will be possible to perform a global fit using simultaneously all sets of data.  At the present stage our approach guarantees a better control over the various kinematical dependences.

\subsection{Flavor Dependence}
\label{sec:flavor}
The $u$ and $d$ quark chiral odd distributions can be readily obtained from Eqs.(\ref{S0}) and (\ref{S1}) 
by using the SU(4) symmetry of the proton wave function,
\begin{eqnarray}
\mid p \uparrow \rangle &=& \sqrt{\frac{2}{1+a_S^2}}  \left[ \frac{a_S}{\sqrt{2}} \mid u\uparrow S_0^0 \rangle + \frac{1}{3 \sqrt{2}} \mid u\uparrow T_0^0 \rangle -\frac{1}{3} \mid u\downarrow T_0^{1} \rangle
-\frac{1}{3} \mid d\uparrow T_{1}^0 \rangle + \frac{\sqrt{2}}{3} \mid d\downarrow T_{1}^{1} \rangle \right]
\end{eqnarray}
where $S_0^0 \equiv S_{I_3}^{S_3}$ is the scalar diquark with isospin 0 and spin component 0, $T_{0,1}^{0,1} \equiv T_{I_3}^{S_3}$ is the axial vector (triplet) diquark with indicated isospin and spin components, and the parameter $a_S=1$ for SU(4) symmetry and can differ from 1 to allow for symmetry breaking \cite{GolLie,BCR}.  Separating out the spin dependence leaves,
\begin{eqnarray}
\mid p \uparrow \rangle &=& \sqrt{\frac{2}{1+a_S^2}}  \left[ \frac{a_S}{\sqrt{2}} \mid u S_0 \rangle \mid 0,\uparrow \rangle + \left( \frac{1}{\sqrt{3}} \mid d \, T_{1} \rangle -\frac{1}{\sqrt{6}} \mid u \,T_0 \rangle \right) \left( \sqrt{\frac{2}{3}} \mid 1, \downarrow \rangle - \sqrt{\frac{1}{3}} \mid 0, \uparrow \rangle \right)
\right] .
\end{eqnarray}
When matrix elements are formed with this state and the corresponding spin down proton, then the sum over the spin states will leave the purely flavor or isospin couplings, 
\begin{equation}
\mid p \rangle =  \sqrt{\frac{2}{1+a_S^2}}  \left[ \frac{a_S}{\sqrt{2}} \mid u \, S_0 \rangle -\frac{1}{\sqrt{6}} \mid u \, T_0 \rangle +  \frac{1}{\sqrt{3}} \mid d \, T_{1} \rangle \right] .
\end{equation}
From here we can see that the spin independent nucleon distributions decompose as 
\begin{eqnarray}
F^u &=& \frac{2}{1+a_S^2} \left( \frac{3}{2}a_S^2 F^{(0)} + \frac{1}{2} F^{(1)} \right) \nonumber \\
\label{Fu}
F^d &=& \frac{2}{1+a_S^2} F^{(1)},
\label{Fd}
\end{eqnarray}
where an overall normalization of $1/3$ for quark number has been imposed and the sum over quark spins has been taken. On the other hand, for the spin transfer GPDs, as in $g_1$ and $h_1$, only the quark spin state $\mid 0,\, \uparrow \rangle$ contributes, so the $F^{(1)}$ is replaced by $-\frac{1}{3}F_T^{(1)}$, 
\begin{eqnarray}
\label{FTu}
F_T^u & = &  \frac{2}{1+a_S^2} \left(\frac{3}{2} a_S^2 F_T^{(0)} - \frac{1}{6} F_T^{(1)}\right)   \nonumber   \\
\label{FTd}
F_T^d & = & -  \frac{2}{1+a_S^2} \frac{1}{3} F_T^{(1)}, 
\end{eqnarray}
where $F_T^{q} \equiv \{ H_T^q, E_T^q, \widetilde{H}_T^q, \widetilde{E}_T^q \}$, $q=u,d$. 
By inverting Eqs.(\ref{Fd},\ref{FTd}), and inserting the expressions for $F^{(0,1)}$, and $F_T^{(0,1)}$ in Eqs.(\ref{Amp0},\ref{Amp1}), we obtain  
the chiral odd GPDs written in terms of their $u$ and $d$ quarks components.

\begin{figure}
\includegraphics[width=9.cm]{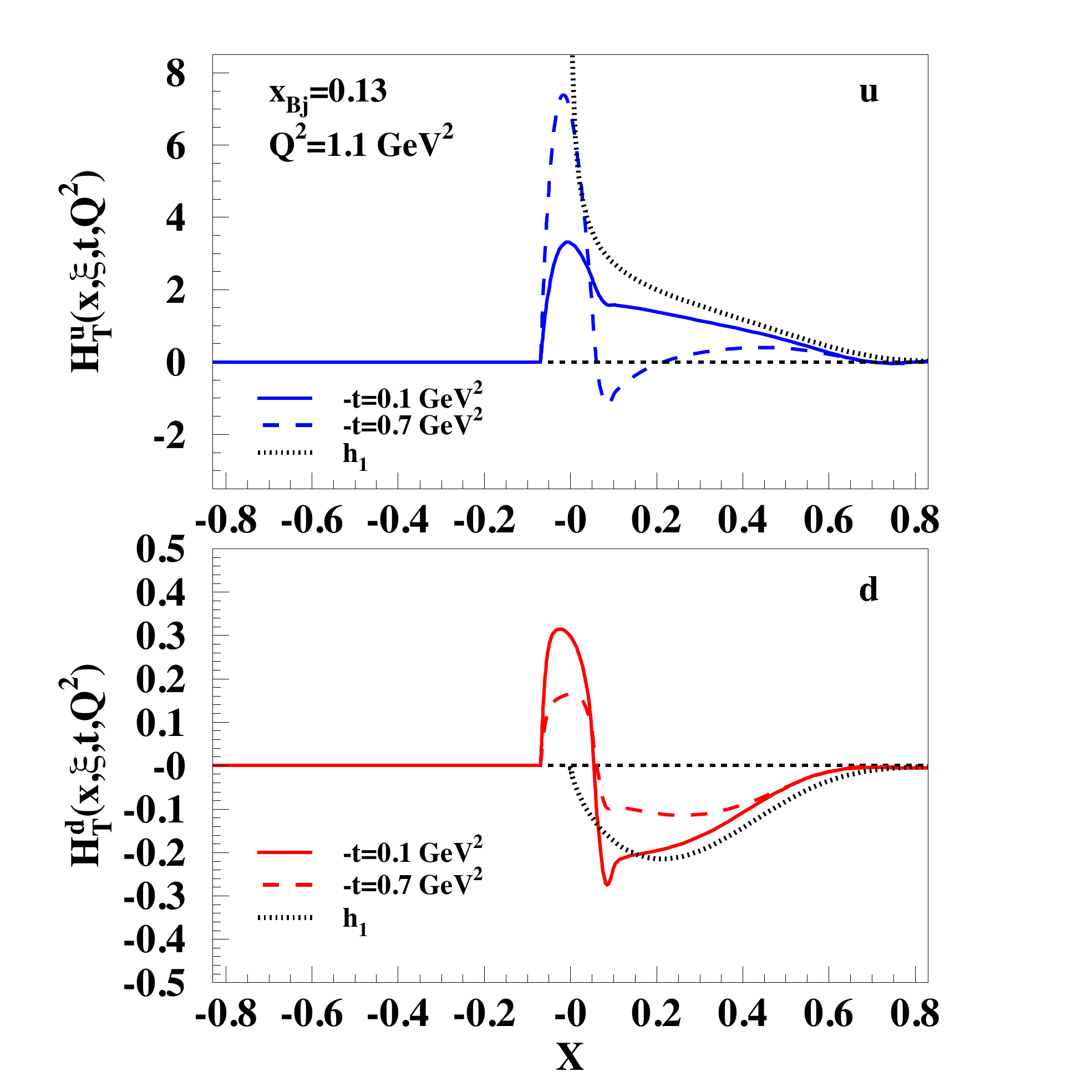}
\caption{(Color online) Chiral odd GPD $H_T^q(x,\xi,t; Q^2)$, for $q=u$ (upper panel) and $q=d$ lower panel. The dotted line is the forward limit, or transversity, $H_T^q(x,0,0;Q^2) \equiv h_1(x,Q^2)$.
All lines were obtained for $x_{Bj}=0.13$ and $Q^2=1.1$ GeV$^2$. The full line is for $-t= 0.1$ GeV$^2$, the dashed line is for $-t=0.7$ GeV$^2$. This GPD dominates the amplitude $f_{10}^{+-}$ at small $-t$.}
\label{gpdht:fig}
\end{figure}
\begin{figure}
\includegraphics[width=8.cm]{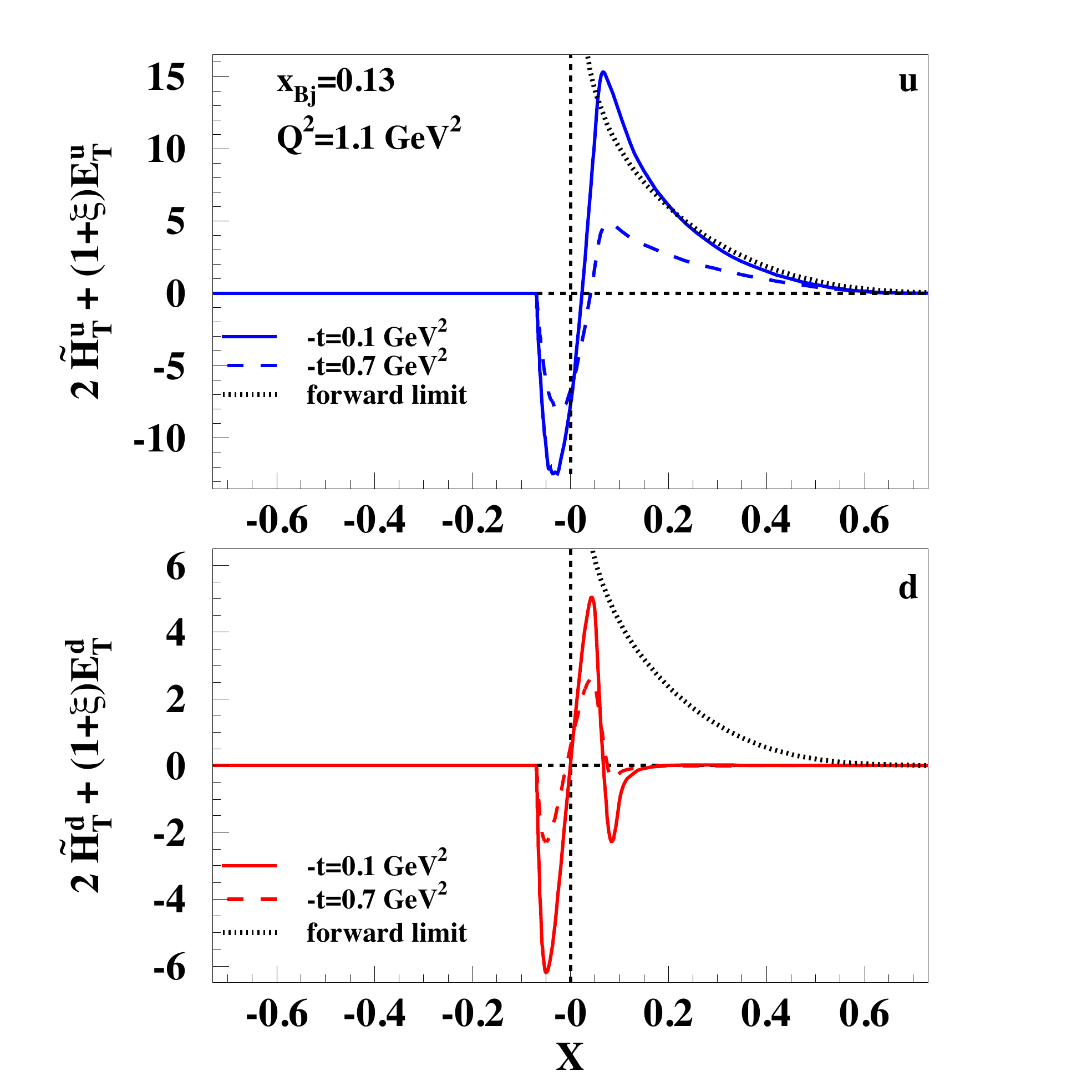}
\includegraphics[width=8.cm]{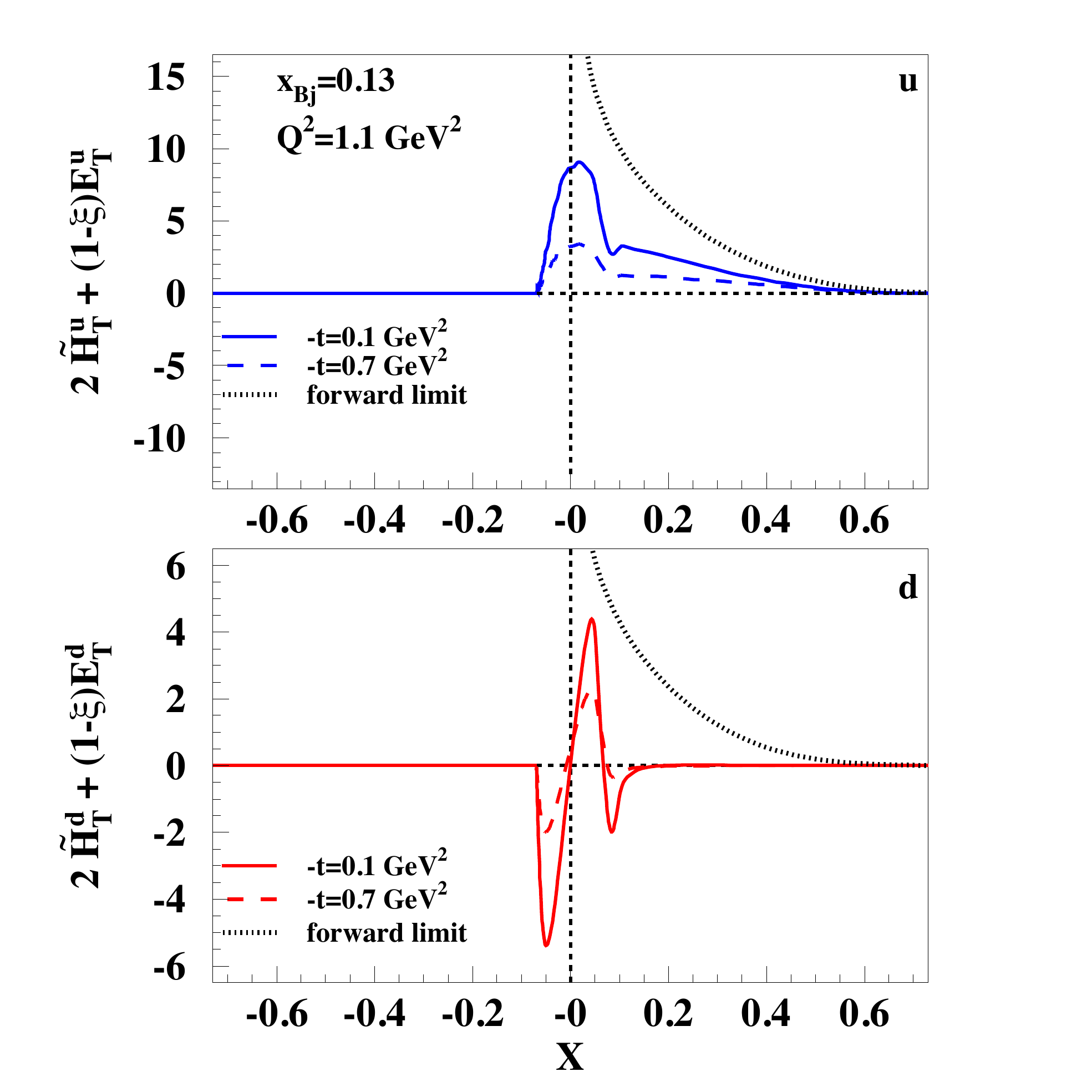}
\caption{(Color online) Chiral odd GPD combinations. Left:  $[2\widetilde{H}_T^q+(1+\xi)E_T^q](x,\xi,t; Q^2)$, for $q=u$ (upper panel) and $q=d$ lower panel, right: $[2\widetilde{H}_T^q+(1-\xi)E_T^q](x,\xi,t; Q^2)$, for $q=u$ (upper panel) and $q=d$ lower panel. The dotted line is the forward limit, or $[2\widetilde{H}_T^q+ E_T^q](x,\xi,t; Q^2)$.
All lines were obtained for $x_{Bj}=0.13$ and $Q^2=1.1$ GeV$^2$. The full line is for $-t= 0.1$ GeV$^2$, the dashed line is for $-t=0.7$ GeV$^2$. These combinations dominate the helicity amplitudes, $f_{10}^{++}$  and $f_{10}^{--}$ , respectively.}
\label{gpdetbar:fig}
\end{figure}
\begin{figure}
\includegraphics[width=9.cm]{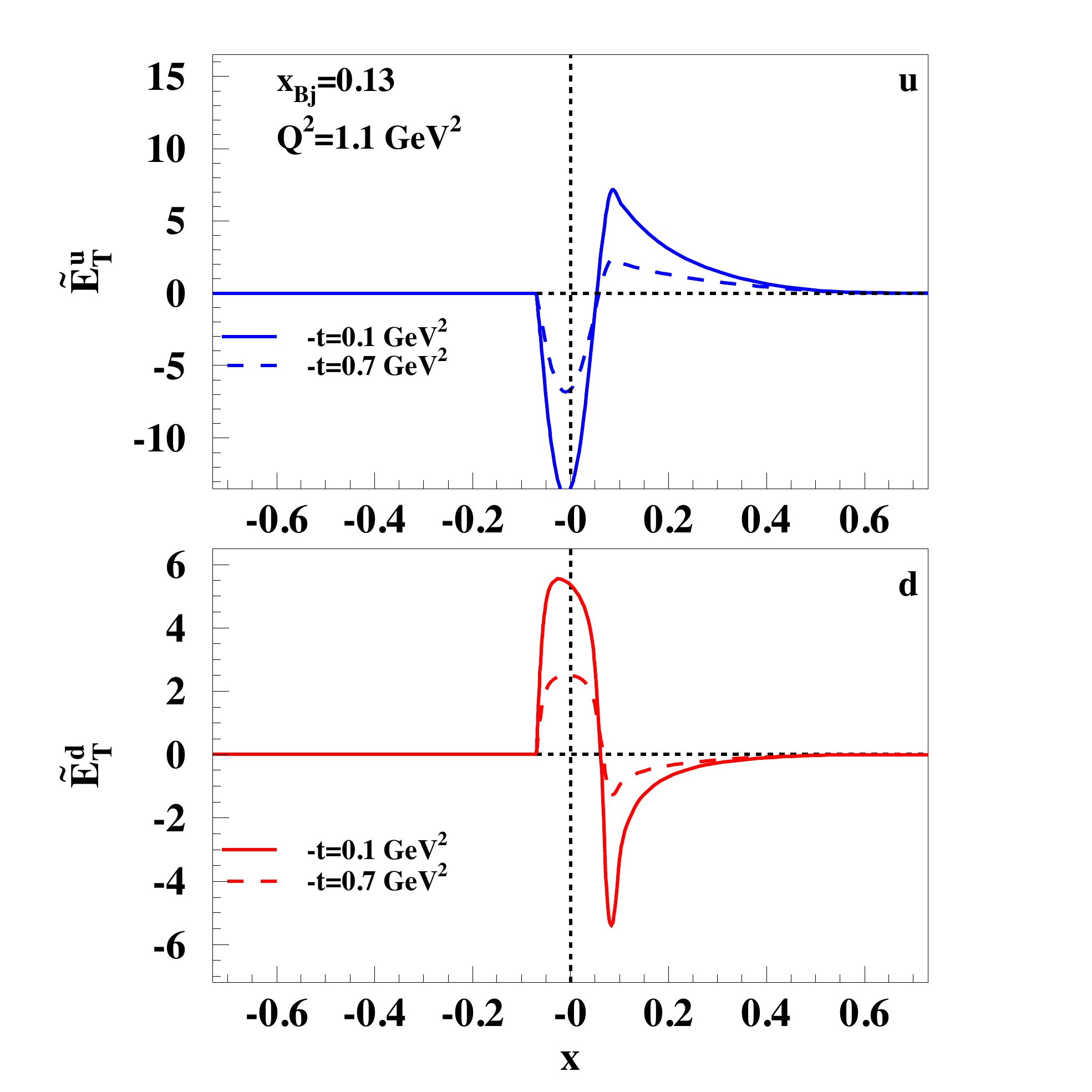}
\caption{(Color online) Chiral odd GPD combinations. $\widetilde{E}_T^q(x,\xi,t; Q^2)$, for $q=u$ (upper panel) and $q=d$ lower panel.
All lines were obtained for $x_{Bj}=0.13$ and $Q^2=1.1$ GeV$^2$. The full line is for $-t= 0.1$ GeV$^2$, the dashed line is for $-t=0.7$ GeV$^2$. $\widetilde{E}_T^q$ enters the helicity amplitudes, $f_{10}^{++}$  and $f_{10}^{--}$, however its contribution is smaller than the GPDs combination in Fig.\ref{gpdetbar:fig}.}
\label{gpdettil:fig}
\end{figure}
In Figures \ref{gpdht:fig}, \ref{gpdetbar:fig}, and \ref{gpdettil:fig} we show the flavor separated GPDs obtained by fixing all parameters of our model using constraints on 
the chiral even GPDs as explained above. Our strategy for extracting chiral odd GPDs from exclusive electroproduction data is  
to gradually loosen such constraints as more data in the chiral odd sector become available.  

To extract flavor dependent  chiral odd GPDs directly from the data it is important to analyze simultaneously $\pi^0$ and $\eta$ production. 
In fact,  from the  SU(3) flavor symmetry for the pseudo-scalar meson octet applied to the chiral odd sector one has,
\begin{subequations}
\label{octet}
\begin{eqnarray}
{\cal F}_T^{\pi^0}  & =  & \frac{1}{\sqrt{2}} (e_u {\cal F}_T^u - e_d {\cal F}_T^d) \\
{\cal F}_T^{\eta}  & =  &  \frac{1}{\sqrt{6}} (e_u {\cal F}_T^u + e_d {\cal F}_T^d - 2 e_s {\cal F}_T^s)
\end{eqnarray}
\end{subequations}
where  $e_q$, $q=u,d,s$, is the quark's charge. 
A flavor separation of the chiral odd CFFs can be performed  by inverting the above equations,  disregarding the contribution of the strange GPD,
\begin{eqnarray}
\label{octet2}
e_u {\cal F}_T^u   & \approx  & \frac{1}{\sqrt{2}}   \left( {\cal F}_T^{\pi^0} + \sqrt{3}  {\cal F}_T^{\eta} \right) \\
-e_d {\cal F}_T^d   & \approx  &   \frac{1}{\sqrt{2}} \left( {\cal F}_T^{\pi^0} - \sqrt{3}  {\cal F}_T^{\eta} \right)
\end{eqnarray}

%

\section{Cross Sections and Asymmetries}
\label{sec:3}
The CFFs and from them the GPDs which were evaluated in the previous Section can be extracted from the cross section terms for exclusive meson electroproduction, which, using the notation of Ref.\cite{Bacchetta_review} (based on \cite{DieSap}),  is defined as, 
\begin{widetext}
\begin{eqnarray} 
\label{xs}
\frac{d^4\sigma}{dx_{Bj} dy d\phi dt} & = & \Gamma \left\{ \left[ F_{UU,T} + \epsilon F_{UU,L}+ \epsilon \cos 2\phi F_{UU}^{\cos 2 \phi} 
+ \sqrt{\epsilon(\epsilon+1)} \cos \phi F_{UU}^{\cos \phi}   + 
h  \, \sqrt{\epsilon(1-\epsilon)} \, \sin \phi F_{LU}^{\sin \phi} \right] \right. \nonumber \\
& + & S_{||} \left[   \sqrt{\epsilon(\epsilon+1)} \sin \phi F_{UL}^{\sin \phi}  + \epsilon \sin 2 \phi F_{UL}^{\sin 2 \phi}  + h \, 
\left( \sqrt{1 - \epsilon^2} \, F_{LL} +    \sqrt{\epsilon(1-\epsilon)} \, \cos \phi \, F_{LL}^{\cos \phi}  \right) \right]   \nonumber \\
%
& - &  S_\perp \left[ \sin(\phi-\phi_S) \left(F_{UT,T}^{\sin(\phi-\phi_S)} + \epsilon  F_{UT,L}^{\sin(\phi-\phi_S)}  \right) + 
\frac{\epsilon}{2} \left( \sin(\phi+\phi_S) F_{UT}^{\sin(\phi+\phi_S)}  +  \sin(3\phi-\phi_S) F_{UT}^{\sin(3\phi-\phi_S)} \right)  \right. \nonumber \\
& + & \left. \sqrt{\epsilon(1+\epsilon)} \left( \sin\phi_S F_{UT}^{\sin \phi_S} + \sin(2\phi-\phi_S) F_{UT}^{\sin(2\phi-\phi_S)} \right) \right]   \nonumber \\
& + & \left. S_\perp h \left[ \sqrt{1-\epsilon^2} \cos(\phi-\phi_S) F_{LT}^{\cos(\phi-\phi_S)} +
\sqrt{\epsilon(1-\epsilon)} \left(\cos \phi_S F_{LT}^{\cos\phi_S} + \cos(2\phi-\phi_S) F_{LT}^{ \cos(2\phi-\phi_S)} \right) \right] \right\} \nonumber \\
\end{eqnarray}
\end{widetext}
where 
$S_{||}$ and ${\bf S}_\perp$ refer to lab frame target polarization parallel and perpendicular to the virtual photon direction, $h$ is the lepton beam helicity, $\phi$ is the azimuthal angle between the lepton plane and the hadron scattering plane, $\phi_S$ is the azimuthal angle of the transverse spin vector ${\bf S}_\perp$ and $t$ is the square of the invariant momentum transfer between the initial and final nucleon.
%

The photon polarization parameter $\epsilon$, the ratio of longitudinal photon and transverse photon flux, can be written in terms of invariants as,
\begin{equation}
\epsilon^{-1} = 1 + 2\left( 1+\frac{\nu^2}{Q^2} \right)\left(4 \dfrac{\nu^2}{Q^2} \dfrac{1-y}{y^2}-1\right)^{-1}. 
\end{equation}
\noindent 
$\Gamma$ is, up to a kinematic factor, given by,
\begin{equation}
\label{Gamma}
\Gamma = \frac{\alpha^2 \, y^2 (1-x_{Bj})}{2\pi x_{Bj}(1-\epsilon)Q^2}.
\end{equation}
%
%
%

In Ref.\cite{GGL_pi0} we analyzed the unpolarized and longitudinally polarized contributions to the cross section, or the various modulations of the $F_{UU}$ and $F_{UL}$, $F_{LL}$, types, respectively.
We showed that: 

\noindent {\it i)}  the polarized cross section terms, $F_{LU},  F_{UL}^{\sin \phi}, F_{LL}^{\cos \phi}$ are dominated by chiral even GPDs through the contribution of longitudinal photon polarization,
while the terms $A_{UL}^{\sin 2\phi}, F_{LL}$ are purely transverse, and therefore useful for the extraction of chiral odd GPDs; 

\noindent {\it ii)} the cross section components with a large chiral odd contribution are dominated by the GPDs $\widetilde{H}_T$, and $E_T$,  and $\widetilde{E}_T$, while $H_T$'s contribution appears only in $F_{LL}$, and it can be disentangled from the other GPDs only at very low $t$ (Figures 15 and 16 in Ref.\cite{GGL_pi0}).

For the single transversely polarized target in Eq.(\ref{xs}), however, there appear terms which are proportional to $H_T$ and which will therefore enable us to extract the tensor charge from data. From Eq.(\ref{xs}) we see that there are  six structure functions for the unpolarized beam and single transversely polarized target,
\begin{eqnarray}
F_{UT,T}^{\sin(\phi-\phi_S)} &=&   \Im m  \, F_{11}^{+-}  =  \Im m  \,  \sum_{\Lambda'}  f_{10}^{+ \Lambda' *}  f_{10}^{- \Lambda'} 
 =  \, \Im m \left[ f_{10}^{++*} f_{10}^{-+} + f_{10}^{+-*} f_{10}^{--} \right] \\
F_{UT,L}^{\sin(\phi-\phi_S)} &=&  \Im m  \, F_{00}^{+-}  = \Im m  \,  \sum_{\Lambda'}  f_{00}^{+ \Lambda' *}  f_{00}^{- \Lambda'}  
 =   \, \Im m \left[ f_{00}^{++*} f_{00}^{-+} + f_{00}^{+-*} f_{00}^{--} \right]   \\
F_{UT}^{\sin(\phi+\phi_S)} &=& \Im m  \, F_{1-1}^{+-}  =  \Im m  \,  \sum_{\Lambda'}  f_{10}^{+ \Lambda' *}  f_{-10}^{- \Lambda'}  
 =  \, \Im m \left[ -f_{10}^{++*} f_{10}^{+-} + f_{10}^{+-*} f_{10}^{++}  \right]  \\
F_{UT}^{\sin(3\phi+\phi_S)} &=& \Im m  \, F_{1-1}^{-+} =  \Im m  \,  \sum_{\Lambda'}  f_{10}^{- \Lambda' *}  f_{-10}^{+ \Lambda'}    
 =  \, \Im m \left[ f_{10}^{-+*} f_{10}^{--} - f_{10}^{--*} f_{10}^{-+}  \right] \\
F_{UT}^{\sin\phi_S} &=& \Im m  \, F_{10}^{+-}  =   \Im m  \,  \sum_{\Lambda'}  f_{10}^{+ \Lambda' *}  f_{00}^{- \Lambda'}   
 =   \, \Im m \left[ f_{10}^{++*} f_{00}^{-+} + f_{10}^{+-*} f_{00}^{--} \right] \\
F_{UT}^{\sin(2\phi-\phi_S)} &=&  \Im m  \, F_{10}^{-+} =  \Im m  \,  \sum_{\Lambda'}  f_{10}^{- \Lambda' *}  f_{00}^{+ \Lambda'}  
 =  \, \Im m \left[ f_{10}^{-+*} f_{00}^{++} + f_{10}^{--*} f_{00}^{+-} \right],
\label{UT}
\end{eqnarray}
and three for the longitudinally polarized lepton and transversely polarized target,
\begin{eqnarray}
F_{LT}^{\cos(\phi-\phi_S)} &=&  \Re e   \, F_{11}^{+-} =  \Re e  \,   \,  \sum_{\Lambda'}  f_{10}^{+ \Lambda' *}  f_{10}^{- \Lambda'} 
= \, \Re e \left[ f_{10}^{++*} f_{10}^{-+} + f_{10}^{+-*} f_{10}^{--} \right]
\label{LT} \\
F_{LT}^{\cos\phi_S} &=&  \Re e   \, F_{10}^{+-}  =  \Re e  \,   \,  \sum_{\Lambda'}  f_{10}^{+ \Lambda' *}  f_{00}^{- \Lambda'} 
= \, \Re e \left[ f_{10}^{++*} f_{00}^{+-} + f_{10}^{+-*} f_{00}^{--} \right] \\
F_{LT}^{\cos(2\phi-\phi_S)} &=&  \Re e   \, F_{10}^{-+}  =    \Re e  \,   \,  \sum_{\Lambda'}  f_{10}^{- \Lambda' *}  f_{00}^{+ \Lambda'} 
=  \, \Re e \left[ f_{10}^{-+*} f_{00}^{++} + f_{10}^{--*} f_{00}^{+-} \right] .
\end{eqnarray}
Notice that: {\it i)} when the nucleon is polarized along the photon direction there will be no asymmetry because of Parity conservation; {\it ii)} when the nucleon is polarized along the incoming lepton direction there will be a component of nucleon polarization transverse to the photon direction as well as transverse to the nucleon plane. This produces  the modulations involving both  the photon angle relative to the lepton beam $\phi_s$, and  the azymuthal angle $\phi$.

Below we list the asymmetries that one can form involving transverse photon polarization only,  so that they are most sensitive to the tensor charge,
\begin{eqnarray}
A_{UT}^{\sin(\phi-\phi_S)} & = & \frac{- F_{UT,T}^{\sin(\phi-\phi_S)}}{F_{UU,T} +\epsilon \, F_{UU,L} } = 
 - \, \frac{(\Re e f_{10}^{++} \Im m f_{10}^{-+} - \Im m f_{10}^{++} \Re e f_{10}^{-+}) + ( \Re e {\bf  f_{10}^{+-}} \Im m f_{10}^{--} - \Im m \, {\bf f_{10}^{+-}} \Re e f_{10}^{--}) }{d\sigma / dt }
\label{A_UTsinphim} \\
A_{UT}^{\sin(\phi+\phi_S)} & = &-\frac{\epsilon}{2} \frac{F_{UT}^{\sin(\phi+\phi_S)}}{F_{UU,T} +\epsilon \, F_{UU,L} } =  - \epsilon \, \frac{ \Re e {\bf f_{10}^{+-}} \,  \Im m f_{10}^{++} - \Re e f_{10}^{++} \,  \Im m {\bf f_{10}^{+-}}}{d \sigma /dt} 
\label{A_UTsinphip} \\
A_{LT}^{\cos(\phi-\phi_S)} & = & \sqrt{1-\epsilon^2} \, \frac{F_{UT}^{\cos(\phi-\phi_S)}}{F_{UU,T} +\epsilon \, F_{UU,L} } \nonumber \\
& = & \sqrt{1-\epsilon^2} \,  \frac{(\Re e f_{10}^{++} \Re e f_{10}^{-+} + \Im m f_{10}^{++} \Im m f_{10}^{-+}) + ( \Re e {\bf  f_{10}^{+-}} \Re e f_{10}^{--} + \Im m \, {\bf f_{10}^{+-}} \Im m  f_{10}^{--}) }{d \sigma/dt}
\label{A_LTcosphi}
\end{eqnarray}
where we have highlighted in boldface the amplitude parts which are sensitive to transversity; the unpolarized cross section is  
\begin{eqnarray}
\label{dsigT}
\frac{d \sigma }{d t} = F_{UU,T} + \epsilon F_{UU,L} 
\end{eqnarray}

Notice that the asymmetry $A_{UT}^{\sin(3\phi-\phi_S)}$, 
\begin{eqnarray}
A_{UT}^{\sin(3\phi-\phi_S)} & = & - \epsilon \, \frac{F_{UT}^{\sin(3\phi-\phi_S)}}{F_{UU,T} +\epsilon \, F_{UU,L} } = -  \epsilon \, \frac{2 \Re e  f_{10}^{--} \,  \Im m f_{10}^{-+} }{d \sigma/dt}, 
\end{eqnarray} 
also involves transverse photon polarization only but it does not involve transversity, and it is predicted to be small being dominated by the double flip amplitude $f_{10}^{-+} $.

\subsection{Results}
\label{results:sec}
\begin{figure}
\includegraphics[width=8.cm]{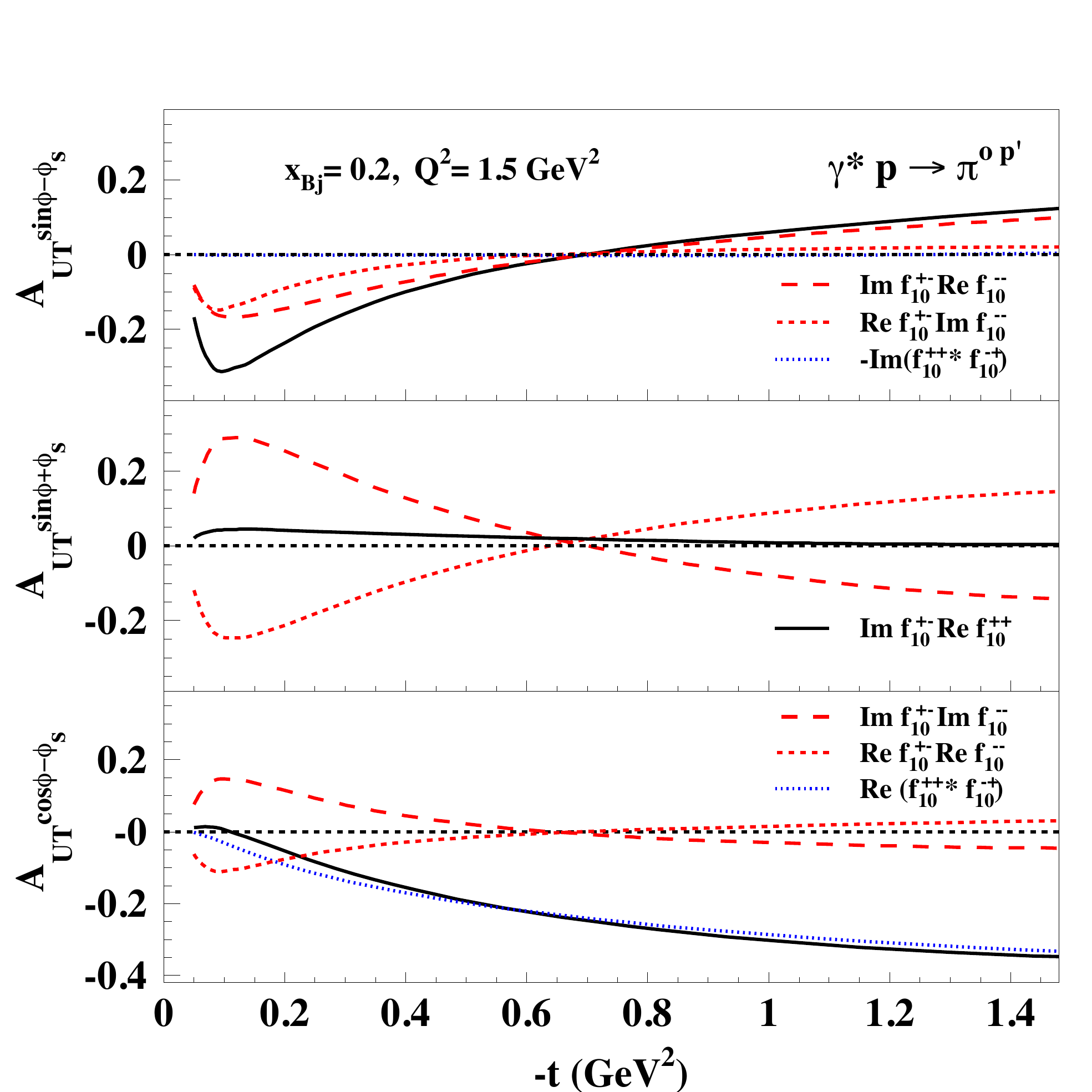}
\includegraphics[width=8.cm]{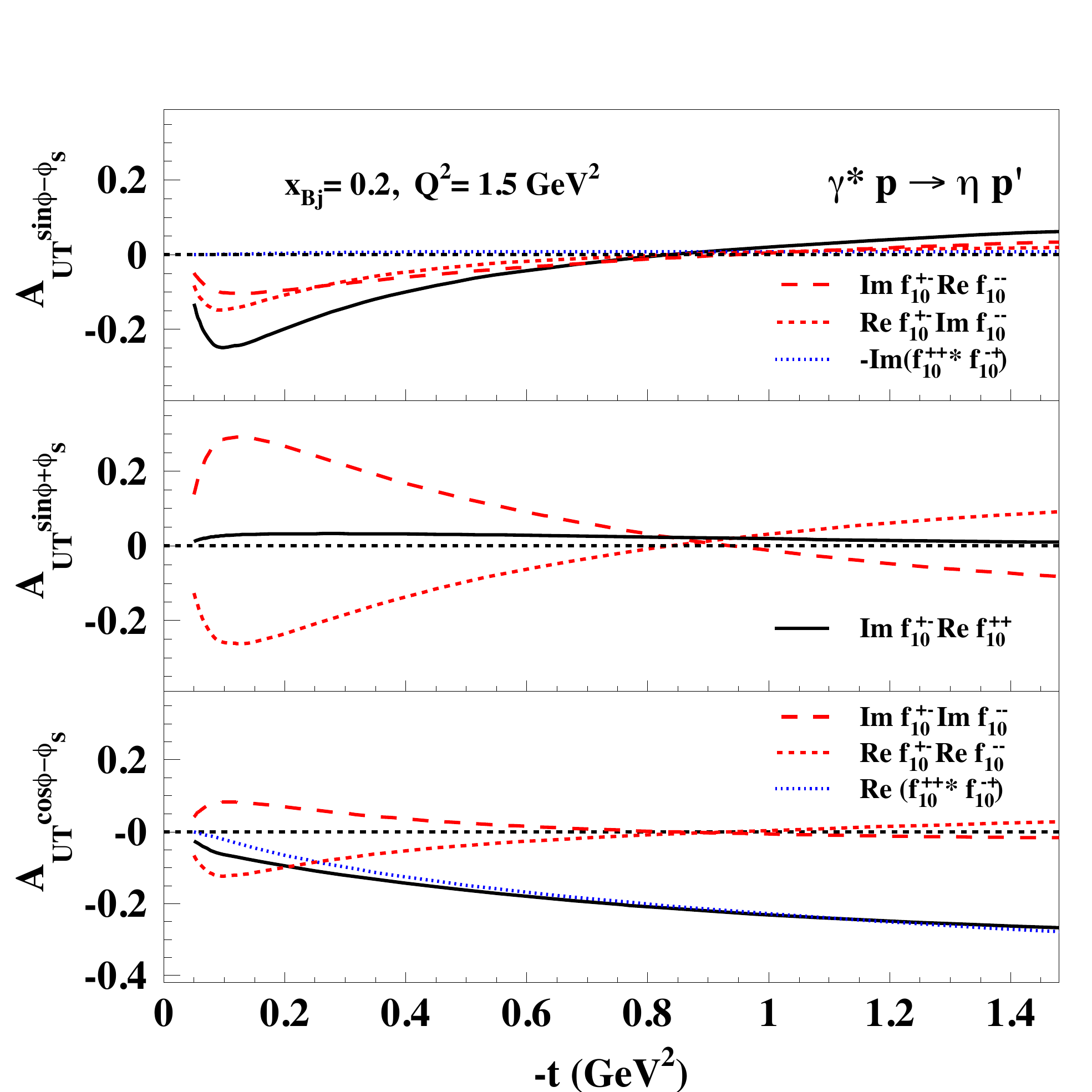}
\caption{(Color online) The asymmetries $A_{UT}^{\sin(\phi-\phi_S)}$, Eq.(\ref{A_UTsinphim}), $A_{UT}^{\sin(\phi+\phi_S)}$, Eq.(\ref{A_UTsinphip}), and $A_{LT}^{\cos(\phi-\phi_S)}$,  Eq.(\ref{A_LTcosphi}) plotted vs. $-t$, at $x_{Bj}=0.2$, $Q^2=1.5$. In the left panel we show $\gamma^* p \rightarrow \pi^0 p'$; in the right panel we show $\gamma^* p \rightarrow \eta p'$. The long dashed and short dashed curves show the contributions proportional to the amplitudes, $\Im m f^{+-}_{10}$ and $\Re e f^{+-}_{10}$, which contain the GPD $H_T$ (Eq.(\ref{helamps_gpd2})), and are therefore sensitive to the tensor charge. The dotted curves show the sum of contributions from the remaining amplitudes, and the full curves show the total contribution. }
\label{pi0aut1}
\end{figure}

We now illustrate the various steps in the procedure for extracting chiral odd GPDs, and their forward limits, in particular transversity, and its integrated value, the tensor charge.
We start by showing in Figure \ref{pi0aut1}, the behavior of $A_{UT}^{\sin(\phi-\phi_S)}$, Eq.(\ref{A_UTsinphim}), $A_{UT}^{\sin(\phi+\phi_S)}$, Eq.(\ref{A_UTsinphip}), and $A_{LT}^{\cos(\phi-\phi_S)}$, Eq.(\ref{A_LTcosphi}),  at kinematics which are attainable at Jefferson Lab, namely $x_{Bj}=0.2$, $Q^2=1.5$. The left panel  shows $\gamma^* p \rightarrow \pi^0 p'$, and the right panel shows $\gamma^* p \rightarrow \eta p'$. The contributions from the terms $\Im m f^{+-}_{10}$ and $\Re e f^{+-}_{10}$ are indicated in the figure by long dashed and short dashed curves, respectively. The other contributions which are not sensitive to the tensor charge but that are sensitive to $\kappa^T_q$ are indicated by the dotted curve. From the figure we deduce that $A_{UT}^{\sin(\phi-\phi_S)}$ is the best quantity to extract the transversity GPD, $H_T$, while the two contributions from the real and imaginary parts of the amps nearly cancel each other in $A_{UT}^{\sin(\phi+\phi_S)}$; $A_{LT}^{\cos(\phi-\phi_S)}$, although it is predicted to be large, is dominated by $f_{10}^{++}$, and therefore it is sensitive to $\kappa_q$.  We also observe clear difference between the $\eta$ and $\pi^0$ curves which can be examined in more detail by considering ratios of the observables for the two processes as we show in what follows.

\begin{figure}
\includegraphics[width=9.cm]{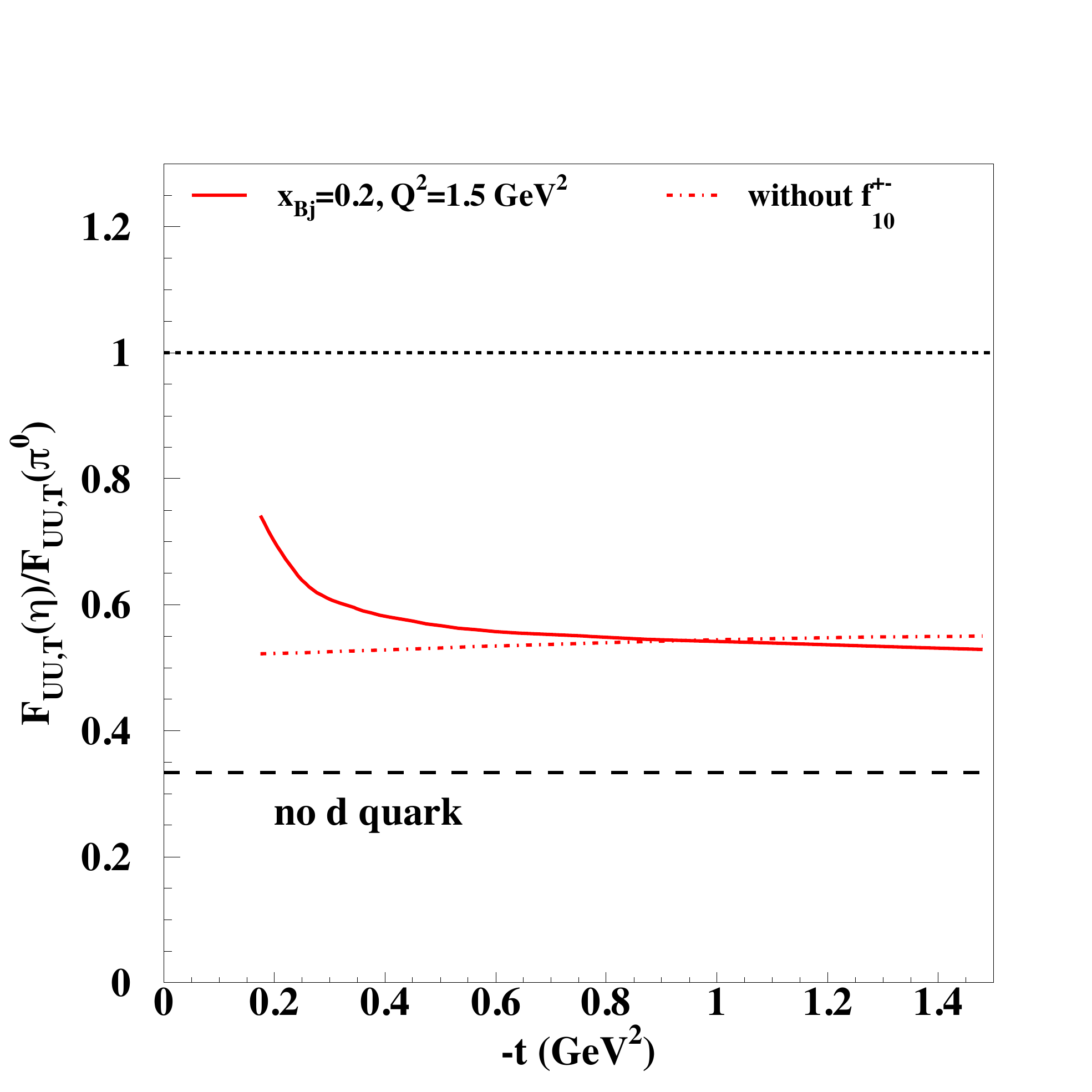}
\caption{((Color online) The ratio of the unpolarized transverse cross section, $F_{UU,T}$ for $\eta$ over $\pi^0$ plotted vs. $-t$ at $x_{Bj}=0.2$ and $Q^2=1.5$ GeV$^2$. 
Because, as explained in the text, the ratio is given by an interplay of the two GPDs $\overline{E}_T$  and $H_T$, in the given  $x_{Bj}$ and $Q^2$ bin, we plot 
both the total contribution (full line), and the ratio obtained omitting the ampltiude $f_{10}^{+-}$ which is dominated by $H_T$  at low $-t$ (dot-dashed line).
Since the ratio would be equal to $1/3$ in the absence of $d$ quark contributions  (Eqs.(\ref{octet}), dashed line), 
its behavior shows the presence of a small but negative  $d$ quark component in both $H_T$ and $\overline{E}_T$. 
}
\label{pi0eta:fig}
\end{figure}

In Figure \ref{pi0eta:fig} we show  the ratio of the unpolarized transverse cross section, $F_{UU,T}$ for $\eta$ over $\pi^0$. This type of plot can be considered a first step towards flavor separation, although many important details should be considered. First of all, the transverse cross section receives contributions from all four transverse helicity amplitudes squared \cite{GGL_pi0}
\[ F_{UU,T} \propto \mid f_{10}^{++} \mid^2 +  \mid f_{10}^{+-} + \mid^2 \mid f_{10}^{-+} + \mid^2 \mid f_{10}^{--} \mid^2 \]
However, the dominant terms are $f_{10}^{++}  \propto \overline{\cal E}_T$, and $f_{10}^{+-} \propto {\cal H}_T$. So each term in the ratio is given by an interplay of the 
two GPDs which are related to the tensor anomalous magnetic moment and to the tensor charge, respectively. For each ($x_{Bj}$, $Q^2$) bin, the $H_T$ term dominates at low $-t$, while the $\overline{E}_T$ term dominates at larger values of $-t$. Now, as one can see from Eqs.(\ref{octet}) the ratio would be equal to $1/3$ in the absence of $d$ quark contributions. 
Therefore, the behavior of the ratio at low $-t$ reflects the sign and magnitude of the $d$ quark contribution to the transversity GPD, $H_T$, while at larger $-t$ it reflects the behavior of the $d$ quarks in $\overline{E}_T$. 

We further clarify this point by showing the ratio without the contribution of $f_{10}^{+-} (H_T)$: one can see a clear difference in the behavior of the two curves at low $-t$. The fact that the curves lie higher than $1/3$ indicates that both $\overline{E}_T^d$ and $H_T^d$ are negative.   As we show later on, the integral of $\overline{E}_T^d$ for $t=0$, $\kappa^T_d$, is positive (see also Fig.\ref{gpdetbar:fig}), however, in the off-forward case  $\overline{E}_T^d$ can oscillate, being negative at $x=\xi$. As a consequence of this behavior, if we consider the integral of $\overline{E}_d$ over $x$, {\it i.e.} the quark tensor anomalous magnetic moment, this will be reduced due to the ranges of negative strength.  Whether this particular picture (and setting of parameters) will be confirmed or not by the data is not of the essence here. What is important is that through combined $\eta$ and $\pi^0$ exclusive electrproduction measurements, and by selecting the appropriate observables as indicated in both this and  our previous analysis \cite{GGL_pi0},  one will be able unravel the partonic structure underlying both the tensor charge and magnetic moment.    
\begin{figure}
\includegraphics[width=8.5cm]{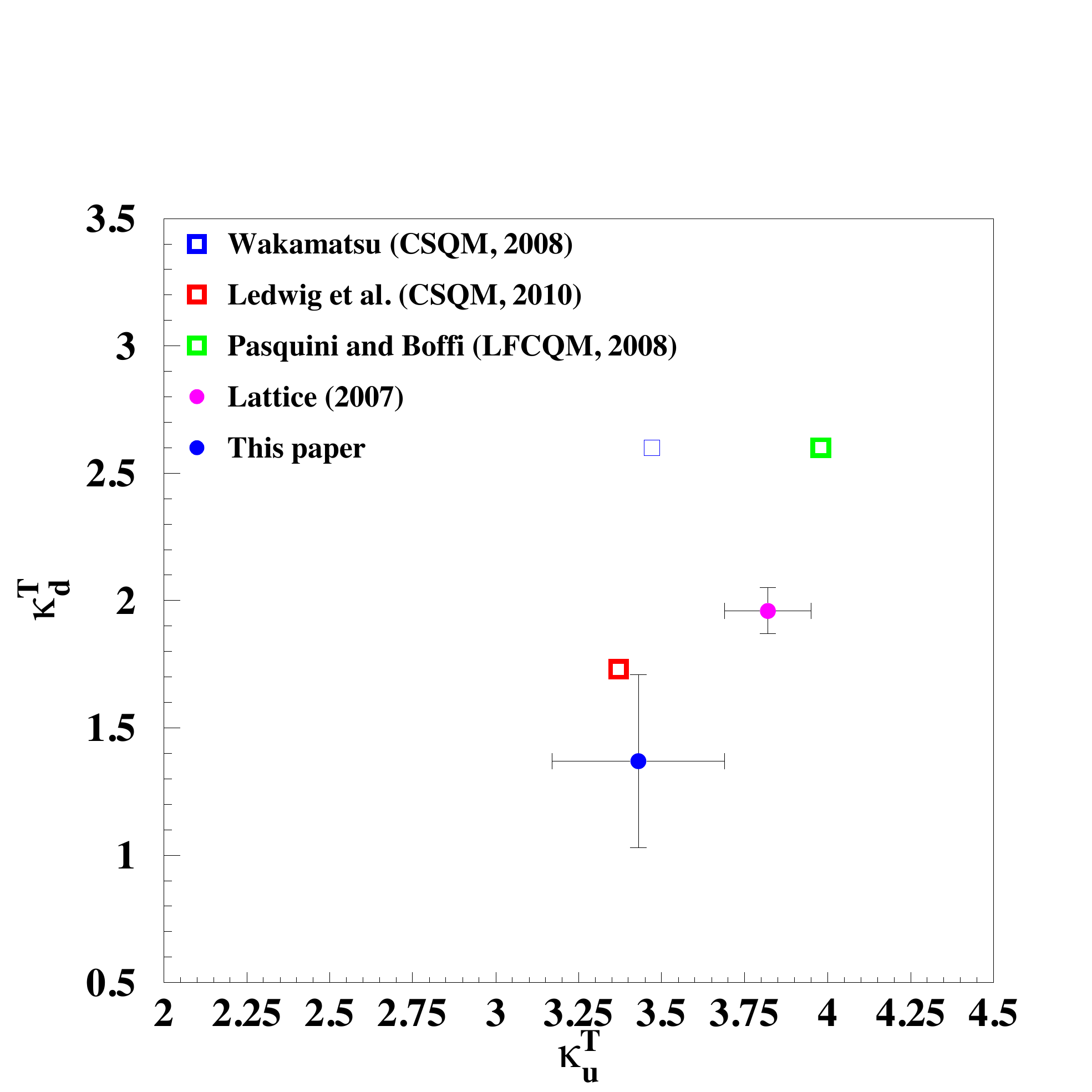}
\caption{(Color online)  The tensor anomalous magnetic moment, $\kappa^T_d$ plotted vs. the $u$ quark, $\kappa^T_u$ extracted in our analysis. 
As for the tensor charge, the error bars on our extraction derive from the extension of our parametrization to the chiral odd sector, {\it i.e.} they are obtained propagating the errors on the chiral even GPDs parameters, as explained in the text. 
Also shown are the values obtained in selected recent models \cite{Waka,PasBof,Ledwig}, and in lattice QCD \cite{lattice}. All values have been evolved to $Q^2=0.8$ GeV$^2$.}
\label{kapu_vs_kapd:fig}
\end{figure}

The tensor anomalous magnetic moment is shown in Figure \ref{kapu_vs_kapd:fig} where the $d$-quark, $\kappa^T_d$ is plotted vs. the value for the $u$ quark, $\kappa^T_u$ (Eq.(\ref{kappaT}). These values were extracted in our analysis simultaneously to the tensor charge. The error bars on our extraction derive from the extension of our parametrization to the chiral odd sector, {\it i.e.} they are obtained propagating the errors on the chiral even GPDs parameters, as explained in the previous Section. To our knowledge, no other determination making use of experimental data has been given so far.
Together with our value, we plot the values obtained in selected recent models \cite{Waka,PasBof,Ledwig}, and the lattice QCD determination from Ref.\cite{lattice}.  All values were calculated by evolving to $Q^2=0.8$ GeV$^2$, noting that the all chiral odd GPDs have  the same anomalous dimensions as for transversity \cite{Ledwig}. The values found in \cite{Waka,PasBof} will produce a ratio $\eta/\pi^0$ much higher than the one plotted in Fig.\ref{pi0eta:fig}. The $\eta/\pi^0$ type of measurements  will allow us, therefore,  to distinguish among models.  
One can see that  $\kappa_q$ is much more undetermined than the tensor charge owing to the scarcity of measurements so far. As we pointed out in Ref.\cite{GGL_pi0}, measurements with a longitudinally polarized  target of both $\pi^0$ and $\eta$ exclusive electroproduction will allow us to  extract the GPD $\overline{E}_T$, and consequently the tensor anomalous magnetic moment.   

\begin{figure}
\includegraphics[width=8.5cm]{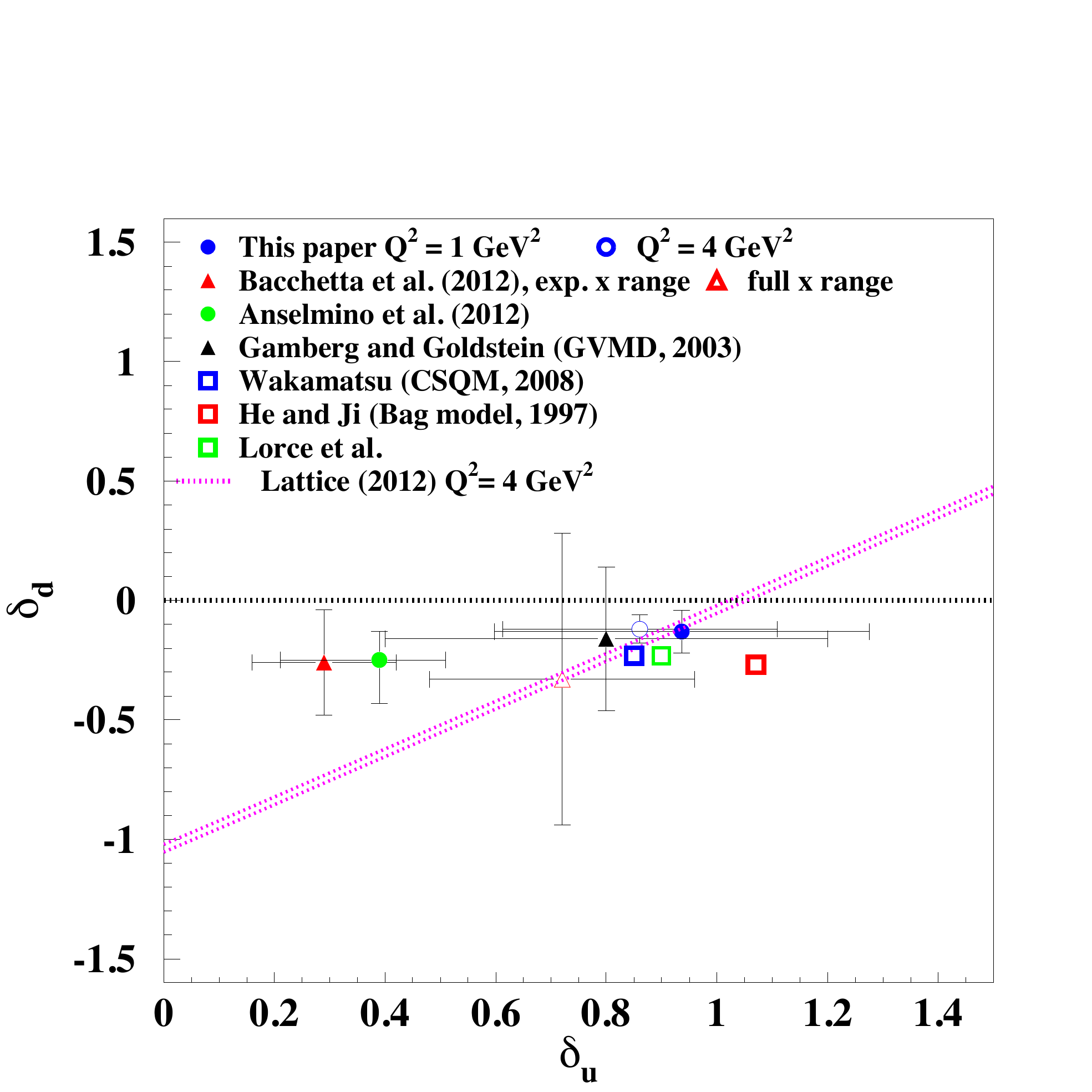}
\caption{(Color online)  Tensor charge values for the $d$ quark, $\delta_d$ plotted vs. the $u$ quark, $\delta_u$, as  obtained from our analysis of exclusive deeply virtual processes, and from other experimental extractions existing to date: dihadron electroproduction production  ($Q^2$= 2 GeV$^2$), Anselmino {\it et al.}, Ref.\cite{Anselmino}, and Bacchetta {\it et al.} Ref.\cite{Courtoy} ($Q^2$= 1 GeV$^2$), and from a model describing the tensor charge through vector and axial vector mesons couplings, Gamberg and Goldstein Ref.\cite{GamGol}  ($Q^2$= 1 GeV$^2$). The thin band delimited by the dotted curves is  the recent lattice QCD result for the isovector component \cite{Engel} ($Q^2$= 4 GeV$^2$). For comparison, the tensor charges obtained in different models are also shown (Wakamatsu, CQSM $Q^2$=0.8 GeV$^2$ Ref.\cite{Waka}, Lorc\'{e} {\it et al.}, CQSM, Ref.\cite{Lorce} ($Q^2$= 1 GeV$^2$), and He and Ji, Bag Model Ref.\protect\cite{HeJi}. For our model we also show the effect of PQCD evolution from $Q^2$=1 GeV$^2$ to $Q^2$= 4 GeV$^2$.}
\label{delu_vs_deld:fig}
\end{figure}

In Figure \ref{delu_vs_deld:fig}  we show along with our results, a compilation of the tensor charge values from recent data analyses besides our suggested one from DV$\pi^0$P and DV$\eta$P, namely the Torino group extraction \cite{Anselmino} obtained combining   data on polarized SIDIS single hadron production \cite{HERMES,COMPASS}, and data on dihadron production from $e^+e^-$ annihilation \cite{Belle}; the Pavia group extraction \cite{Courtoy} obtained from dihadron production in a collinear framework, {\it i.e.} combining the ($k_T$ integrated) transversity distribution, $h_1$ with dihadron fragmentation functions; and finally, what can be considered a pioneering extraction using a combination of vector and axial vector  meson couplings to the nucleon which are constrained from data on the mesons decay constants and the average parton transverse momenta \cite{GamGol}. For comparison we show also the most recent lattice results obtained for the isovector combination $\delta u - \delta d$ \cite{Engel}, and selected model calculations Refs.\cite{HeJi,Waka,Lorce}. The error bars in our extraction are the uncorrelated errors from the parameters in the chiral even sector. 

The values we obtained with the variant of the Reggeized diquark model based fit used in  this paper for both $\delta_q$ and $\kappa_q^T$ are shown in Table \ref{tensor:table}.

\begin{table}
\begin{tabular}{|c|c|c|c|}
\hline
 & $u$ & $d$ & $Q^2$ (GeV$^2$) \\
 \hline
 $\delta_q$ & $0.936 \pm 0.339$ & $-0.130 \pm 0.089$ & 1 \\
 \hline
  $\delta_q$ & $0.860 \pm 0.248 $ & $-0.119 \pm 0.060  $ & 4  \\
 \hline
 $\kappa^T_q$ & $3.43 \pm 0.26 $ & $1.37 \pm 0.34 $ & 0.8 \\
\hline
\end{tabular}
\label{tensor:table}
\caption{Values of the tensor charge and the tensor anomalous magnetic moment obtained in our analysis.}
\end{table}

\begin{figure}
\includegraphics[width=8.cm]{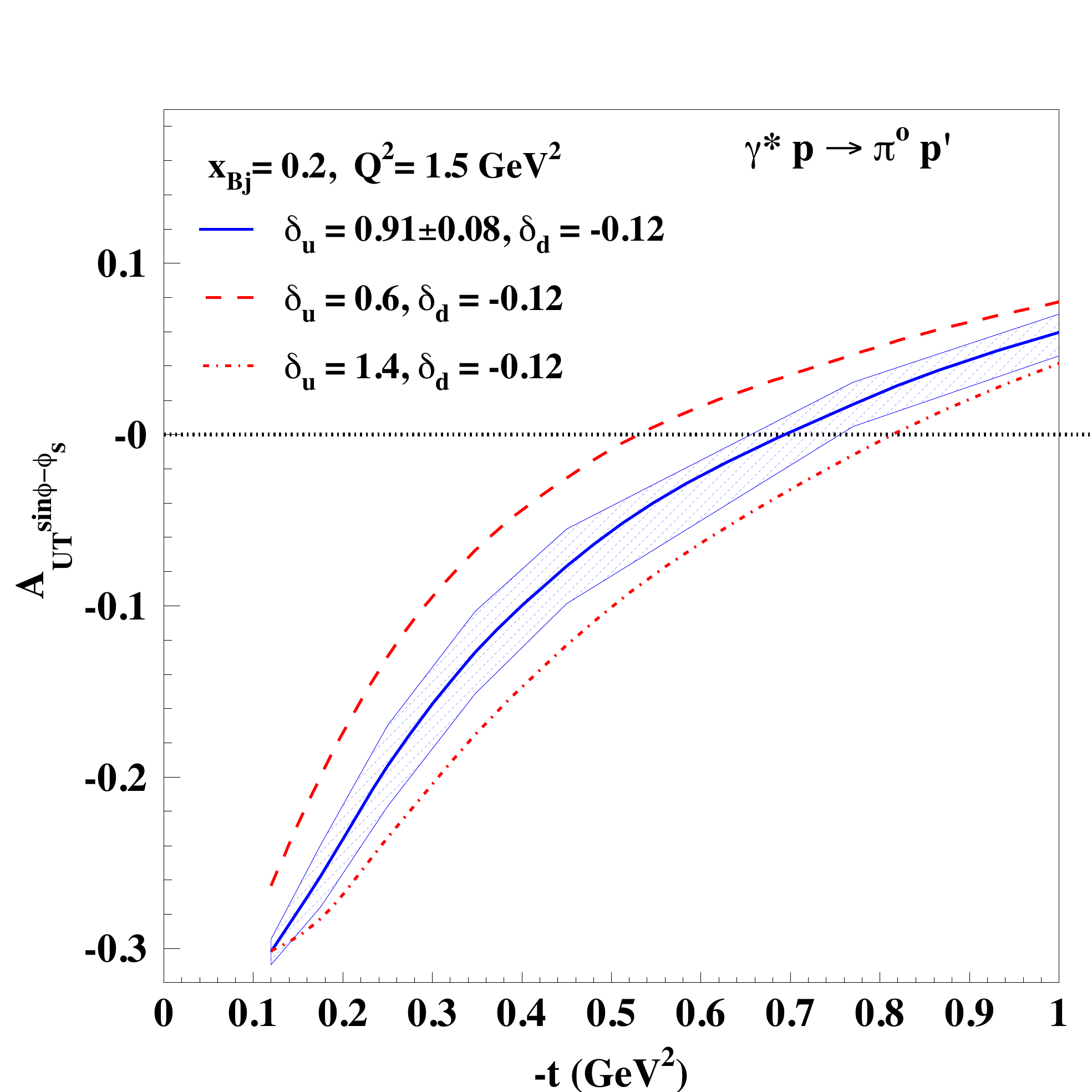}
\caption{(Color online) The asymmetry $A_{UT}^{\sin(\phi-\phi_S)}$, Eq.(\ref{A_UTsinphim}), plotted vs. $-t$, at $x_{Bj}=0.2$, $Q^2=1.5$ for the $\gamma^* p \rightarrow \pi^0 p'$ reaction. The error band was obtained by varying the value of the $u$-quark tensor charge, $\delta_u$, by $\pm 0.08$. The dot-dashed curve corresponds to $\delta_u = 1.4$, and the dashed curve corresponds to $\delta_u=0.6$. The value of $\delta_d$ was kept fixed at $-0.12$. The graph shows the sensitivity of the asymmetry to variations of the tensor charge, or the precision that is needed in measurements of this quantity in order to reduce the size of the errors from the ones reported in Fig.\ref{delu_vs_deld:fig}.}
\label{AUT_tensor}
\end{figure}

Several remarks are in order. 

First of all,  the tensor charge is subject to a rather rapid Perturbative QCD evolution (see \cite{Waka} and references therein), as can be seen from the shift (decrease) in values from $Q^2$=1 GeV$^2$ to $Q^2$= 4 GeV$^2$ shown for our extracted values (all the other evaluations shown in the figure are in the $Q^2$ range: $0.8 - 2$ GeV$^2$).
\begin{figure}
\includegraphics[width=8.5cm]{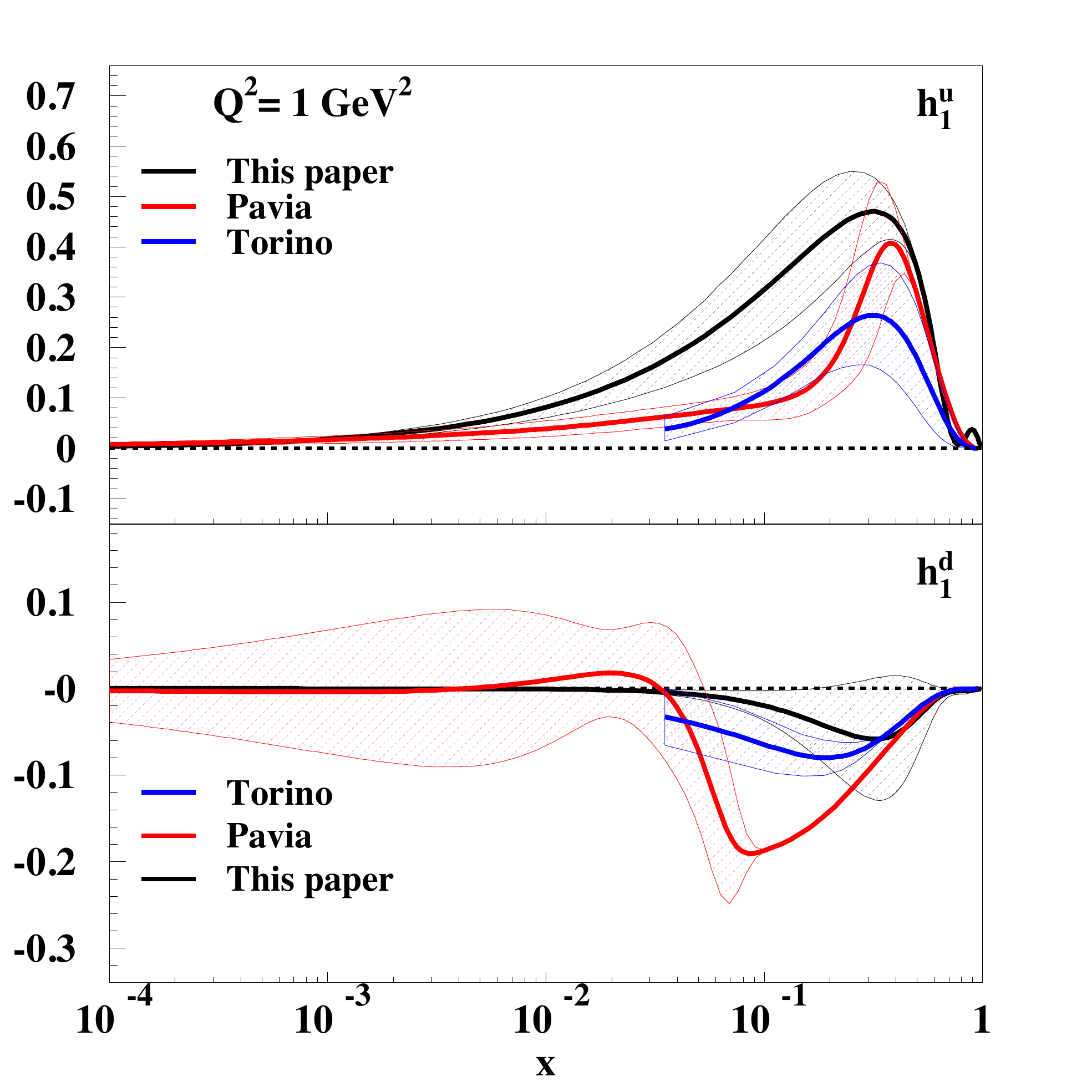}
\caption{(Color online)  Transversity, $h_1^q$ plotted vs. $x$ at $Q^2$= 1 GeV$^2$, for the $u$ quarks (upper panel) and for the $d$ quark (lower panel). Besides our analysis, the most recent extractions from experimental data were plotted, namely the analysis of the Pavia group extraction \cite{Courtoy} obtained from dihadron production in a collinear framework, and the Torino group extraction \cite{Anselmino} obtained combining   data on polarized SIDIS single hadron production \cite{HERMES,COMPASS}, and data on dihadron production from $e^+e^-$ annihilation \cite{Belle} (see the corresponding tensor charge values in Fig.\ref{delu_vs_deld:fig}).}
\label{transv:fig}
\end{figure}

Most importantly, the graph shows how the low $x$ tail of the transversity distribution plays a fundamental role in determining the value of the charge. This can be seen by comparing the values obtained for the $u$ quarks ($d$ quarks values tend to be smaller and the effects are less visible). In fact the values from Ref.\cite{Courtoy} were obtained by calculating the charge in the $x_{Bj}$ range of the fitted data ($x_{Bj} \gtrsim 0.06$), and by extrapolating to $x=10^{-5}$ (open symbol).  The extrapolated result tends to agree better with our extraction. Note that our parametrization explicitly includes Regge behavior, therefore our value of the charge tends to be larger. Our Regge behavior, in turn, is constrained by the nucleon's form factors behavior \cite{newFF}. This is why measuring the tensor charge through GPDs provides a definite advantage over all of the inclusive measurements where extrapolation procedures in the low $x$ regime need to be devised. 

In Figure \ref{AUT_tensor} we demonstrate the sensitivity of pseudoscalar meson production data to the values of the tensor charge. For illustration, we show only the $\pi^0$ asymmetry, and we fix $\delta_d = -0.12$. The figure shows curves for varying  $\delta_u$ in the range $0.6 -1.4$, whereas the error band was obtained by varying the tensor charge by $\pm 0.08$.
Additional information on $\delta_d$ can be obtained by comparing $\pi^0$ and $\eta$ measurements.  The purpose of the calculations represented in this figure is to gain information on the kind of precision that will be needed in order to reduce the error on the tensor charge with respect to the values from the analyses reported in Fig.\ref{delu_vs_deld:fig}. From the figure we conclude that an accurate extraction of tensor charge and its flavor dependence will be possible at Jefferson Lab, 12 GeV \cite{AvaKim}.

The various transversity distributions mentioned in our discussion above are shown in Figure \ref{transv:fig}. These curves show that the tensor charge for the u-quark will be larger than the results from the two other analyses, while the d-quark tensor charge will be smaller, again showing the importance of the small $x$ behavior in our Reggeized scheme.

We conclude this section by noting that the GPD $\widetilde{E}_T$ enters the transverse non flip amplitudes, $f_{10}^{++(--)}$ and it can therefore be extracted once more accurate data are available. We postpone the discussion of this observable to future work.
  
\section{Conclusions and Outlook}
\label{sec:conclusions}
In conclusion, we reiterate that transverse asymmetries allow us to single out the tensor charge best. Namely, the asymmetries are sensitive to the GPD $H_T$, in contrast to the longitudinal target polarization asymmetries reported in detail in Ref.\cite{GGL_pi0} that  are mostly sensitive to the GPDs $\widetilde{H}_T$, $E_T$, and $\widetilde{E}_T$, and to the integral of $2\widetilde{H}_T$ + $E_T$, or the tensor anomalous magnetic moment. A combined analysis of $\pi^0$ and $\eta$ data will allow us to perform a flavor separation of both the tensor charges and the tensor anomalous magnetic moments. 

It should be noticed, however, that although the agreement of the tensor charge values reported in this paper with the very precise recent lattice results is excellent, 
in our analysis the tensor charge was obtained indirectly, by using constraints from Parity relations that allow us to connect to the somewhat better constrained GPDs in the chiral even sector. The extraction we proposed can therefore be considered  model dependent. Nevertheless, at this stage our study provides, on one side, a set of very much needed estimates and constraints on the size of the various, so far largely unexplored, chiral odd GPDs. On the other side, it opens the way to an upcoming analysis that will be performed by fitting our functional forms directly to the combined exclusive $\eta$ and $\pi^0$ electroproduction data, once these will be made available. 
Our ultimate goal is to determine the chiral odd GPDs from a global analysis on its own merit, using all of the pseudoscalar meson production data.  Hence, this paper can be considered to be a step in this direction in that it provides
a framework  with which to gauge the various contributions to all cross sections and asymmetries. 

Most importantly, the suggested analysis will allow us  to substantially reduce the errors on the flavor dependent $\delta_q$, and to perform, for the first time, an experimental extraction of $\kappa_q$.  

We complete our discussion by acknowledging similar work in this direction {\it i.e.} Refs.\cite{GolKro,Kro_new}, and the alternative method proposed in Ref. \cite{Pire1,Pire2} to access chiral odd GPDs, through the electroproduction of two vector mesons.

\acknowledgments
We thank  the Hall B collaboration at Jefferson Lab, in particular Harut Avakian, Francois Xavier Girod, Andrey Kim, Valery Kubarovsky and Paul Stoler for useful discussions and suggestions. We also thank Aurore Courtoy and Alexei Prokudin for discussions and for providing the calculated transversity functions from their respective collaborations' papers. This work was supported by the U.S. Department
of Energy grant DE-FG02-01ER4120.

\appendix

\section{Quark-proton scattering helicity amplitudes}
\label{appa}
The vertex functions corresponding to Fig.\ref{fig_LCWF} are, 
\begin{equation}
 \phi_{-\Lambda -\lambda} = (-1)^{\Lambda-\lambda} \phi^*_{\Lambda \lambda}. 
 \label{Parity0}
 \end{equation}
 where $  \phi^*_{\Lambda \lambda}$, the basic structures  in our model, are the quark-proton scattering amplitudes at leading order with proton-quark-diquark vertices given in Fig.\ref{fig_LCWF}.
The intermediate diquark system can have $J^P=0^+$ (scalar), or $J^P=1^+$ (axial vector). Its invariant mass, $M_X$ varies in our model according to a spectral function, thus generating Regge behavior at large $M_X^2$ \cite{BroCloGun}. We start from the region $X\geq \zeta$. At fixed $M_X$, the amplitudes read,
\begin{equation} 
\label{AmpS0}
A^{(0)}_{\Lambda^\prime \lambda^\prime, \Lambda \lambda}
  =  \int d^2k_\perp\phi^*_{\Lambda^\prime \lambda^\prime}(k^\prime,P^\prime) \phi_{\Lambda \lambda}(k,P),
\end{equation} 
with vertex structures
\begin{subequations}
\begin{eqnarray}
& \phi^*_{++}(k,P) & =  {\cal A}  \left(m+M X \right),           \\
& \phi^*_{+-}(k,P) & =  {\cal A}  ({k}_1 + i {k}_2),        \\   
& \phi_{--}(k,P)   &  =      \phi_{++}(k,P)    \\
&  \phi_{-+}(k,P)&  =  -\phi^*_{+-}(k,P).
\end{eqnarray}
\end{subequations}
For $S=1$,
\begin{equation} 
\label{AmpS1}
A^{(1)}_{\Lambda^\prime \lambda^\prime, \Lambda \lambda}
  =  \int d^2k_\perp\phi^{*\mu}_{\Lambda^\prime \lambda^\prime} (k^\prime,P^\prime) \sum_{\lambda^{\prime \prime}}  \epsilon_\mu^{* \, \lambda^{\prime \prime}}  \epsilon_\nu^{\lambda^{\prime \prime}}  \phi_{\Lambda,\lambda}^\nu (k,P),
\end{equation} 
where $\lambda^{\prime \prime}$ is the diquark's helicity, which in our model is taken as transverse only, and  
the vertex structures are
\begin{subequations}
 \begin{eqnarray}
& \phi_{++}^+(k,P) = &    {\cal A} \,  \frac{k_1 - i k_2}{1-X}    \\
& \phi_{++}^-(k,P) = &   -{\cal A} \,  \frac{(k_1 + i k_2)X}{1-X}  \\
& \phi_{+-}^+(k,P) = &   0   \\
& \phi_{+-}^-(k,P) = &  -{\cal A} \left(m+M X\right)  \\
& \phi_{-+}^+(k,P) = &  -{\cal A} \left(m+M X\right)  \\
& \phi_{-+}^-(k,P) = & 0 
\end{eqnarray}
\end{subequations}
where, 
\begin{equation}
{\cal A} \equiv {\cal A}(k) = \frac{1}{\sqrt{X}} \frac{\Gamma(k)}{k^2-m^2}, \;\;\; {\rm and}  \;\;\; k^2-m^2  =  X M^2 - \frac{X}{1-X} M_X^2 - m^2  - \frac{k_\perp^2}{1-X} .
\end{equation}
For $(k,P) \rightarrow (k',P')$, $X \rightarrow X^\prime = (X-\zeta)/(1-\zeta)$ and $k_i \rightarrow \tilde{k}_i = k_i - (1-X)/(1-\zeta) \Delta_i$, $(i=1,2)$.


\end{document}